\pdfoutput=1

\documentclass[twocolumn,epjc3]{svjour3}

\newcommand{\comm}[1]{}

\RequirePackage[T1]{fontenc}

\smartqed

\RequirePackage{graphicx}
\RequirePackage{amssymb}
\RequirePackage{epsf}
\RequirePackage{psfig}
\RequirePackage{mathptmx}
\RequirePackage{flushend}
\RequirePackage[numbers]{natbib}
\RequirePackage{hyperref}

\journalname{Nonlinear Dynamics}

\begin{document}

\title{Escape dynamics and fractal basin boundaries in Seyfert galaxies}

\author{Euaggelos E. Zotos}

\institute{Department of Physics, School of Science, \\
Aristotle University of Thessaloniki, \\
GR-541 24, Thessaloniki, Greece \\
Corresponding author's email: {evzotos@physics.auth.gr}}

\date{Received: 18 August 2014 / Accepted: 8 January 2015 / Published online: 30 January 2015}

\titlerunning{Escape dynamics and fractal basin boundaries in Seyfert galaxies}

\authorrunning{Euaggelos E. Zotos}

\maketitle

\begin{abstract}

The escape dynamics in a simple analytical gravitational model which describes the motion of stars in a Seyfert galaxy is investigated in detail. We conduct a thorough numerical analysis distinguishing between regular and chaotic orbits as well as between trapped and escaping orbits, considering only unbounded motion for several energy levels. In order to distinguish safely and with certainty between ordered and chaotic motion, we apply the Smaller ALingment Index (SALI) method. It is of particular interest to locate the escape basins through the openings around the collinear Lagrangian points $L_1$ and $L_2$ and relate them with the corresponding spatial distribution of the escape times of the orbits. Our exploration takes place both in the physical $(x,y)$ and in the phase $(x,\dot{x})$ space in order to elucidate the escape process as well as the overall orbital properties of the galactic system. Our numerical analysis reveals the strong dependence of the properties of the considered escape basins with the total orbital energy, with a remarkable presence of fractal basin boundaries along all the escape regimes. It was also observed, that for energy levels close to the critical escape energy the escape rates of orbits are large, while for much higher values of energy most of the orbits have low escape periods or they escape immediately to infinity. We also present evidence obtained through numerical simulations that our model can \comm{realistically} describe the formation and the evolution of the observed spiral structure in Seyfert galaxies. We hope our outcomes to be useful for a further understanding of the escape mechanism in galaxies with active nuclei.


\end{abstract}

\section{Introduction}
\label{intro}

Over the last decades a huge amount of research work has been devoted on the subject of escaping particles from open dynamical systems. Especially the issue of escape in Hamiltonian systems is a classical problem in nonlinear dynamics (e.g., [\citealp{C90}, \citealp{CK92} -- \citealp{CHLG12}, \citealp{STN02}]). It is well known, that several types of Hamiltonian systems have a finite energy of escape. For values of energy lower than the escape energy the equipotential surfaces of the systems are close which means that orbits are bound and therefore escape is impossible. For energy levels above the escape energy on the other hand, the equipotential surfaces open and exit channels emerge through which the particles can escape to infinity. The literature is replete with studies of such ``open" Hamiltonian systems (e.g., [\citealp{BBS09}, \citealp{KSCD99}, \citealp{NH01}, \citealp{STN02}, \citealp{SCK95} -- \citealp{SKCD96}, \citealp{Z13}, \citealp{Z14}]). At this point we should emphasize, that all the above-mentioned references on escapes in Hamiltonian system are exemplary rather than exhaustive, taking into account that a vast quantity of related literature exists.

Nevertheless, the issue of escaping orbits in Hamiltonian systems is by far less explored than the closely related problem of chaotic scattering. In this situation, a test particle coming from infinity approaches and then scatters off a complex potential. This phenomenon is well investigated as well interpreted from the viewpoint of chaos theory (e.g., [\citealp{BGOB88} -- \citealp{BOG89}, \citealp{BM92}, \citealp{CDO90}, \citealp{DGO90} -- \citealp{EJ86}, \citealp{GR89}, \citealp{H88}, \citealp{JRS92}, \citealp{J87} -- \citealp{JT91}, \citealp{LMG00} -- \citealp{LJ99}, \citealp{ML02}, \citealp{RJ94}, \citealp{SASL06} -- \citealp{SS10}]). Chaotic scattering has also been applied in the astrophysical context of many aspects such as scattering off black holes (e.g., [\citealp{A97}, \citealp{dML99}]) and three-body stellar systems (e.g., [\citealp{BTS96}, \citealp{BST98}, \citealp{H83}, \citealp{HB83}]). The related invariant manifolds of the chaotic saddle is directly associated with the chaotic dynamical behavior (e.g., [\citealp{O02}]). In particular, the chaotic saddle is defined as the intersection of its stable and unstable manifolds [\citealp{S14}], while hyperbolic and non-hyperbolic chaotic saddles may occur in dynamical systems (e.g., [\citealp{LGB93}]). More details on the issue of chaotic scattering and escape from chaotic systems can be found to the recent reviews [\citealp{SS13}] and [\citealp{APT13}], respectively.

In open Hamiltonian systems an issue of paramount importance is the determination of the basins of escape, similar to basins of attraction in dissipative systems or even the Newton-Raphson fractal structures. An escape basin is defined as a local set of initial conditions of orbits for which the test particles escape through a certain exit in the equipotential surface for energies above the escape value. Basins of escape have been studied in many earlier papers (e.g., [\citealp{BGOB88}, \citealp{C02}, \citealp{KY91}, \citealp{PCOG96}]). The reader can find more details regarding basins of escape in [\citealp{C02}], while the review [\citealp{Z14ip}] provides information about the escape properties of orbits in a multi-channel dynamical system of a two-dimensional perturbed harmonic oscillator. The boundaries of an escape basins may be fractal (e.g., [\citealp{AVS09}, \citealp{BGOB88}]) or even respect the more restrictive Wada property (e.g., [\citealp{AVS01}]), in the case where three or more escape channels coexist in the equipotential surface.

One of the most characteristic Hamiltonian systems of two degrees of freedom with escape channels is undoubtedly the well-known H\'{e}non-Heiles system [\citealp{HH64}]. A huge load of research on the escape properties of this system has been conducted over the years (e.g., [\citealp{AVS01} -- \citealp{AVS03}, \citealp{BBS08}, \citealp{BBS09}]). Escaping orbits in the classical Restricted-Three-Body-Problem (RTBP) is another typical example (e.g., [\citealp{N04}, \citealp{N05}, \citealp{TdA14}]). Furthermore, escaping and trapped motion of stars in stellar systems is an another issue of great importance. In a recent article [\citealp{Z12a}], we explored the nature of orbits of stars in a galactic-type potential, which can be considered to describe local motion in the meridional $(R,z)$ plane near the central parts of an axially symmetric galaxy. It was observed, that apart from the trapped orbits there are two types of escaping orbits, those which escape fast and those which need to spend vast time intervals inside the equipotential surface before they find the exit and eventually escape. Furthermore, the chaotic dynamics within a star cluster embedded in the tidal field of a galaxy was explored in [\citealp{EJS08}]. In particular, by scanning thoroughly the phase space and obtaining the basins of escape with the respective escape times it was revealed, that the higher escape times correspond to initial conditions of orbits near the fractal basin boundaries.

The aim of this work is to numerically investigate the escape dynamics in a Seyfert galaxy with a rotating active central nucleus. The recent paper of Ernst \& Peters (2014) [\citealp{EP14}] is used as a guide since we follow the same numerical methods and computational approach. The present paper however, contains several new aspects with respect to [\citealp{EP14}] that can be considered as further developments in the field of open galactic systems. In particular: (i) we proceed one step further by using the SALI method in order to distinguish between regular and chaotic trapped motion, (ii) we display the evolution of the percentages of all types of orbits in both the physical and phase space as a function of the value of the energy integral, (iii) we conduct an overview analysis thus revealing how the orbital structure of the dynamical system is influenced by the total energy when we use a continuous spectrum of energy values rather than some few discrete levels and (iv) we relate the escape times of orbits with the energy as well as with the width of the symmetrical escape channels.

The article is organized as follows: In Section \ref{mod} we present in detail the structure and the properties of our Seyfert galaxy model. All the computational methods we used in order to determine the character of orbits are described in Section \ref{cometh}. In the following Section, we conduct a thorough numerical investigation revealing the overall orbital structure (bounded regions and basins of escape) of the galaxy and how it is affected by the total orbital energy. In Section \ref{spr} we show that our dynamical model can \comm{sufficiently} simulate, in away, the creation as well as the evolution of spiral arms of a real Seyfert galaxy. Our paper ends with Section \ref{disc}, where the discussion and the conclusions of this research are given.

\section{Theory of the galaxy model}
\label{mod}

For the description of the Seyfert galaxy we use the following mass-model potential
\begin{equation}
\Phi_{\rm g}(x,y) = \frac{- \ G \ M_{\rm g}}{\sqrt{x^2 + b y^2 + c^2}},
\label{pot}
\end{equation}
where $G$ is the gravitational constant, $M_{\rm g}$ is the total mass of the galaxy, $b$ is a parameter leading to deviation from axial and spherical symmetry, while $c$ is the scale length of the galaxy which acts also as a softening parameter. The generalized anharmonic Plummer potential (\ref{pot}) has been successfully applied several times in the past in order to realistically model non axially symmetric galaxies (see e.g., [\citealp{CP07}, \citealp{MAB95}, \citealp{PC06}, \citealp{Z12b}]). At this point we would like to stress out that the single-component potential of Eq. (\ref{pot}) models the properties of the Seyfert galaxy as a whole rather than an isolate component (e.g., the galactic bar).

We consider the case where the dense and compact active nucleus of the Seyfert galaxy rotates clockwise at a constant angular velocity $\Omega > 0$. Therefore, the effective potential is given by
\begin{equation}
\Phi_{\rm eff}(x,y) = \Phi_{\rm g}(x,y) - \frac{\Omega^2}{2} \left(x^2 + y^2 \right),
\label{eff}
\end{equation}
which is in fact a cut through the equatorial plane of the three-dimensional (3D) potential.

In our study, we use a system of galactic units, where the unit of length is 1 kpc, the unit of mass is $2.325 \times 10^7 {\rm M}_\odot$ and the unit of time is $0.9778 \times 10^8$ yr (approximately 100 Myr). The velocity unit is 10 km/s, while $G$ is equal to unity $(G = 1)$. The energy unit (per unit mass) is 100 km$^2$s$^{-2}$. In these units, the values of the involved parameters are: $M_{\rm g} = 5000$ (corresponding to 1.1625 $\times 10^{11}$ ${\rm M}_\odot$), $b = 2.5$, $c = 1.2$, $\Omega = 1.25$ and they remain constant throughout the numerical analysis. The values of $M_{\rm g}$ and $c$ were chosen having in mind the Seyfert galaxy NGC 1566 (see at the end of section \ref{spr} for more details).

\begin{figure*}[!tH]
\centering
\resizebox{0.9\hsize}{!}{\includegraphics{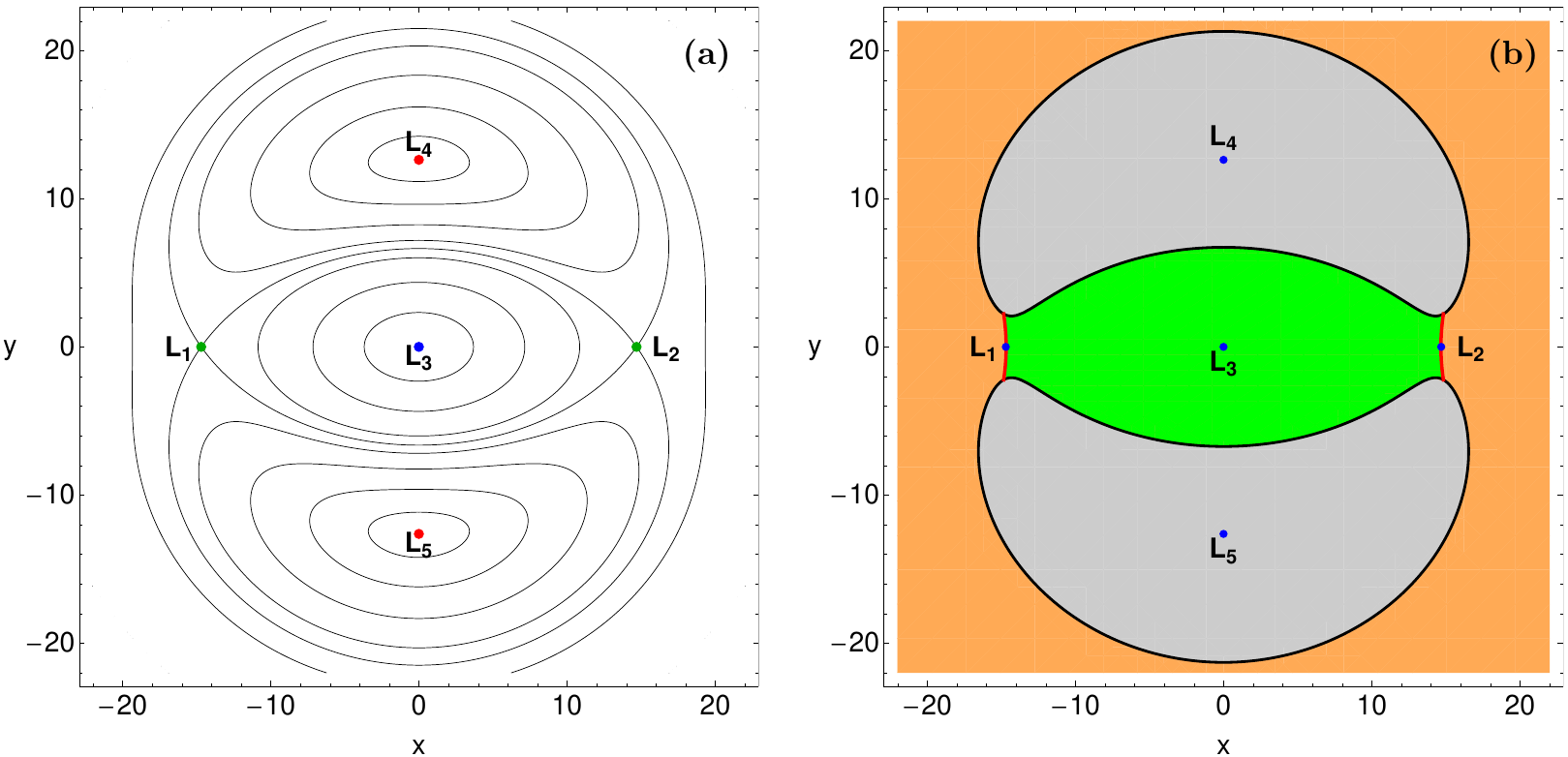}}
\caption{(a-left): The isolines contours of the constant effective potential and the position of the five Lagrangian points. (b-right): The contour of the effective potential $\Phi_{\rm eff}(x,y) = E$ when $\widehat{C} = 0.01$. The escaping orbits leak out through exit channels 1 and 2 of the equipotential surface passing either $L_1$ or $L_2$, respectively. The interior galactic region is indicated in green, the exterior region is shown in amber color, while the forbidden regions of motion are marked with grey. The two unstable Lyapunov orbits around the Lagrangian points are visualized in red.}
\label{conts}
\end{figure*}

This galactic model has five equilibria called Lagrangian points [\citealp{S67}] at which
\begin{equation}
\frac{\partial \Phi_{\rm eff}}{\partial x} = \frac{\partial \Phi_{\rm eff}}{\partial y} = 0.
\label{lps}
\end{equation}
The isolines contours of constant effective potential as well as the position of the five Lagrangian points $L_i, \ i = {1,5}$ are shown in Fig. \ref{conts}a. Three of them, $L_1$, $L_2$, and $L_3$, known as the collinear points, are located in the $x$-axis. The central stationary point $L_3$ at $(x,y) = (0,0)$ is a local minimum of $\Phi_{\rm eff}$. At the other four Lagrangian points it is possible for the test particle to move in a circular orbit, while appearing to be stationary in the rotating frame. For this circular orbit, the centrifugal and the gravitational forces precisely balance. The stationary points $L_1$ and $L_2$ at $(x,y) = (\pm r_L,0) = (\pm 14.687185207778671,0)$ are saddle points, where $r_L$ is called Lagrangian radius. Let $L_1$ located at $x = -r_L$, while $L_2$ be at $x = +r_L$. The points $L_4$ and $L_5$ on the other hand, which are located at $(x,y) = (0, \pm 12.626321712993062)$ are local maxima of the effective potential, enclosed by the banana-shaped isolines. The annulus bounded by the circles through $L_1$, $L_2$ and $L_4$, $L_5$ is known as the ``region of coroation" (see also [\citealp{BT08}]).

The equations of motion in the rotating frame are described in vectorial form by
\begin{equation}
\ddot{{\bf{r}}} = - {\bf{\nabla}} \Phi_{\rm eff} - 2\left(\Omega \times \dot{{\bf{r}}} \right),
\label{eqmot0}
\end{equation}
where $\vec{r} = (x,y,z)$ is the position vector, $\Omega = (0,0,\Omega)$ is the constant rotation velocity vector around the vertical $z$-axis, while the term $- 2\left(\Omega \times \dot{\vec{r}} \right)$ represents the Coriolis force. Decomposing Eq. (\ref{eqmot0}) into rectangular Cartesian coordinates $(x,y)$, we obtain
\begin{eqnarray}
\ddot{x} &=& - \frac{\partial \Phi_{\rm eff}}{\partial x} + 2\Omega\dot{y}, \nonumber \\
\ddot{y} &=& - \frac{\partial \Phi_{\rm eff}}{\partial y} - 2\Omega\dot{x},
\label{eqmot}
\end{eqnarray}
where the dot indicates derivative with respect to the time.

In the same vein, the equations describing the evolution of a deviation vector ${\bf{w}} = (\delta x, \delta y, \delta \dot{x}, \delta \dot{y})$ which joins the corresponding phase space points of two initially nearby orbits, needed for the calculation of standard chaos indicators (the SALI in our case) are given by the following variational equations
\begin{eqnarray}
\dot{(\delta x)} &=& \delta \dot{x}, \nonumber \\
\dot{(\delta y)} &=& \delta \dot{y}, \nonumber \\
(\dot{\delta \dot{x}}) &=&
- \frac{\partial^2 \Phi_{\rm eff}}{\partial x^2} \ \delta x
- \frac{\partial^2 \Phi_{\rm eff}}{\partial x \partial y} \ \delta y + 2\Omega \ \delta \dot{y},
\nonumber \\
(\dot{\delta \dot{y}}) &=&
- \frac{\partial^2 \Phi_{\rm eff}}{\partial y \partial x} \ \delta x
- \frac{\partial^2 \Phi_{\rm eff}}{\partial y^2} \ \delta y - 2\Omega  \ \delta \dot{x}.
\label{vareq}
\end{eqnarray}

Consequently, the corresponding Hamiltonian (known also as the Jacobian) to the effective potential given in Eq. (\ref{eff}) reads
\begin{equation}
H_{\rm J}(x,y,\dot{x},\dot{y}) = \frac{1}{2} \left(\dot{x}^2 + \dot{y}^2 \right) + \Phi_{\rm eff}(x,y) = E,
\label{ham}
\end{equation}
where $\dot{x}$ and $\dot{y}$ are the momenta per unit mass, conjugate to $x$ and $y$, respectively, while $E$ is the numerical value of the Jacobian, which is conserved since it is an isolating integral of motion. Thus, an orbit with a given value for it's energy integral is restricted in its motion to regions in which $E \leq \Phi_{\rm eff}$, while all other regions are forbidden to the star. The numerical value of the effective potential at the two Lagrangian points $L_1$ and $L_2$, that is $\Phi_{\rm eff}(-r_L,0)$ and $\Phi_{\rm eff}(r_L,0)$, respectively yields to a critical Jacobi constant $C_L = -507.828303111454$, which can be used to define a dimensionless energy parameter as
\begin{equation}
\widehat{C} = \frac{C_L - E}{C_L},
\label{chat}
\end{equation}
where $E$ is some other value of the Jacobian. The dimensionless energy parameter $\widehat{C}$ makes the reference to energy levels more convenient. For $E = C_L$ the equipotential surface encloses the critical area\footnote{In three dimensions the last closed equipotential surface of the effective potential passing through the Lagrangian points $L_1$ and $L_2$ encloses a critical volume.}, while for a Jacobian value $E > C_L$, or in other words when $\widehat{C} > 0$, the equipotential surface is open and consequently stars can escape from the galaxy. In Fig. \ref{conts}b we present the contour of $\Phi_{\rm eff}(x,y) = E$ when $\widehat{C} = 0.01$. We observe, the two openings (exit channels) through which the particles can leak out. In fact, we may say that these two exits act as hoses connecting the interior region (green color) of the galaxy where $-r_L < x < r_L$ with the ``outside world" of the exterior region (amber color). Exit channel 1 at the negative $x$-direction corresponds to escape through Lagrangian point $L_1$, while exit channel 2 at the positive $x$-direction indicates escape through $L_2$. The forbidden regions of motion within the banana-shaped isolines around $L_4$ and $L_5$ points are shown in Fig. \ref{conts}b with gray. It is evident, that for $E < C_L$ the two forbidden regions merge together thus escape is impossible since the interior region cannon communicate with the exterior area.

\section{Computational methods}
\label{cometh}

In order to explore the escape dynamics of stars in the Seyfert galaxy model, we need to define samples of initial conditions of orbits whose properties (bounded or escaping motion) will be identified. The best method for this purpose, would have been to choose the sets of initial conditions of the orbits from a distribution function of the system. This however, is not available, so we define for each value of the total orbital energy (all tested energy levels are above the escape energy), dense grids of initial conditions regularly distributed in the area allowed by the value of the energy. Our investigation takes place both in the physical $(x,y)$ and in the phase $(x,\dot{x})$ space for a better understanding of the escape process. In both cases, the step separation of the initial conditions along the axes (or in other words, the density of the grids) is controlled in such a way that always there are about $10^5$ orbits (a maximum grid of 318 $\times$ 318 equally spaced initial conditions of orbits), depending on the value of the Jacobian integral. Furthermore, the grids of initial conditions of orbits whose properties will be examined are defined as follows: For the physical $(x,y)$ space we consider orbits with initial conditions $(x_0, y_0)$ with $\dot{y_0} = 0$, while the initial value of $\dot{x_0}$ is always obtained from the energy integral (\ref{ham}) as $\dot{x_0} = \dot{x}(x_0,y_0,\dot{y_0},E) > 0$. Similarly, for the phase $(x,\dot{x})$ space we consider orbits with initial conditions $(x_0, \dot{x_0})$ with $y_0 = 0$, while this time the initial value of $\dot{y_0}$ is obtained from the Jacobi integral (\ref{ham}).

The equations of motion as well as the variational equations for the initial conditions of all orbits were integrated using a double precision Bulirsch-Stoer \verb!FORTRAN 77! algorithm (e.g., [\citealp{PTVF92}]) with a small time step of order of $10^{-2}$, which is sufficient enough for the desired accuracy of our computations. Here we should emphasize, that our previous numerical experience suggests that the Bulirsch-Stoer integrator is both faster and more accurate than a double precision Runge-Kutta-Fehlberg algorithm of order 7 with Cash-Karp coefficients. Throughout all our computations, the Jacobian energy integral (Eq. (\ref{ham})) was conserved better than one part in $10^{-11}$, although for most orbits it was better than one part in $10^{-12}$.

In dynamical systems with escapes an issue of paramount importance is the determination of the position as well as the time at which an orbit escapes. For all values of energy smaller than the critical value $C_L$ (escape energy), the equipotential surface is closed and therefore all orbits are bounded. On the other hand, when $E > C_L$ the equipotential surface is open and extend to infinity. The value of the energy itself however, does not furnish a sufficient condition for escape. An open equipotential curve consists of two branches forming channels through which an orbit can escape to infinity (see Fig. \ref{conts}b). It was proved [\citealp{C79}] that in Hamiltonian systems there is a highly unstable periodic orbit at every opening close to the line of maximum potential which is called a Lyapunov orbit. In fact, there is a family of unstable Lyapunov orbits around each collinear Lagrangian point [\citealp{M58}]. Such an orbit reaches the Zero Velocity Curve\footnote{The boundaries of the accessible regions of motion are the called Zero Velocity Curves, since they are the locus in the $(x,y)$ space where kinetic energy vanishes.} (ZVC), on both sides of the opening and returns along the same path thus, connecting two opposite branches of the ZVC. Lyapunov orbits are very important for the escapes from the system, since if an orbit intersects any one of these orbits with velocity pointing outwards moves always outwards and eventually escapes from the system without any further intersections with the surface of section (e.g., [\citealp{C90}]. When $E = C_L$ the Lagrangian points exist precisely but when $E > C_L$ an unstable Lyapunov periodic orbit is located close to each of these two points (e.g., [\citealp{H69}]). The two unstable Lyapunov orbits around $L_1$ and $L_2$ are shown in Fig. \ref{conts}b in red.

\begin{figure}[!tH]
\includegraphics[width=\hsize]{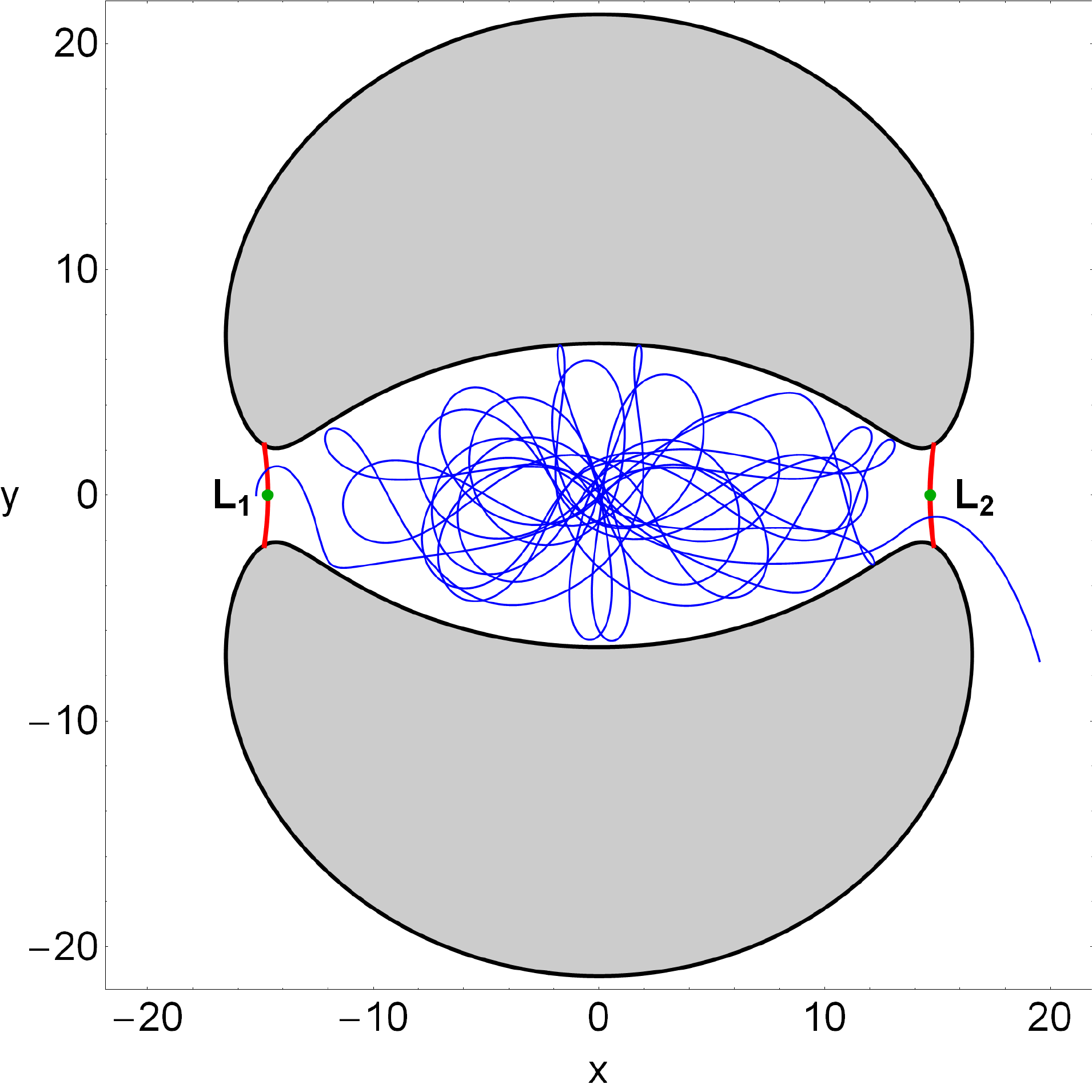}
\caption{An orbit with initial conditions outside the interior region which however does not escape immediately from the system. In particular, even though the orbit is initiated outside but very close to $L_1$ it eventually escapes from the opposite exit channel, through $L_2$.}
\label{orb12}
\end{figure}

These two Lagrangian points are saddle points of the effective potential, so when $E > C_L$, a star must pass close enough to one of these points in order to escape. Thus, in our galactic system the escape criterion is purely geometric. In particular, escapers are defined to be those stars moving in orbits beyond the Lagrangian radius $(r_L)$ thus passing one of the two Lagrangian points ($L_1$ or $L_2$) with velocity pointing outwards. Here we must emphasize that orbits with initial conditions outside the interior region, or in other words outside $L_1$ or $L_2$, does not necessarily escape immediately from the galaxy. In Fig. \ref{orb12} we present a characteristic example of such an orbit with initial conditions: $x_0 = -15.18 > r_L$, $y_0 = \dot{x_0} = 0$, $\dot{y_0} > 0$ when $\widehat{C} = 0.01$. We observe that even though this orbit is initiated outside but very close to $L_1$ it does not escape right away from the system. On the other hand, it enters the interior galactic region and only after 32 time units of chaotic motion it eventually escapes from the opposite channel. Thus it is evident, that the initial position itself does not furnish a sufficient condition for escape, since the escape criterion is a combination of the coordinates and the velocity of stars.

The physical and the phase space are divided into the escaping and non-escaping (trapped) space. Usually, the vast majority of the trapped space is occupied by initial conditions of regular orbits forming stability islands where a third integral is present. In many systems however, trapped chaotic orbits have also been observed. Therefore, we decided to distinguish between regular and chaotic trapped motion. Over the years, several chaos indicators have been developed in order to determine the character of orbits. In our case, we chose to use the Smaller ALingment Index (SALI) method. The SALI [\citealp{S01}] has been proved a very fast, reliable and effective tool, which is defined as
\begin{equation}
\rm SALI(t) \equiv min(d_-, d_+),
\label{sali}
\end{equation}
where $d_- \equiv \| {\bf{w_1}}(t) - {\bf{w_2}}(t) \|$ and $d_+ \equiv \| {\bf{w_1}}(t) + {\bf{w_2}}(t) \|$ are the alignments indices, while ${\bf{w_1}}(t)$ and ${\bf{w_2}}(t)$, are two deviation vectors which initially point in two random directions. For distinguishing between ordered and chaotic motion, all we have to do is to compute the SALI along time interval $t_{max}$ of numerical integration. In particular, we track simultaneously the time-evolution of the main orbit itself as well as the two deviation vectors ${\bf{w_1}}(t)$ and ${\bf{w_2}}(t)$ in order to compute the SALI.

The time-evolution of SALI strongly depends on the nature of the computed orbit since when an orbit is regular the SALI exhibits small fluctuations around non zero values, while on the other hand, in the case of chaotic orbits the SALI after a small transient period it tends exponentially to zero approaching the limit of the accuracy of the computer $(10^{-16})$. Therefore, the particular time-evolution of the SALI allow us to distinguish fast and safely between regular and chaotic motion. Nevertheless, we have to define a specific numerical threshold value for determining the transition from order to chaos. After conducting extensive numerical experiments, integrating many sets of orbits, we conclude that a safe threshold value for the SALI is the value $10^{-8}$. Thus, in order to decide whether an orbit is regular or chaotic, one may follow the usual method according to which we check after a certain and predefined time interval of numerical integration, if the value of SALI has become less than the established threshold value. Therefore, if SALI $\leq 10^{-8}$ the orbit is chaotic, while if SALI $ > 10^{-8}$ the orbit is regular thus making the distinction between regular and chaotic motion clear and beyond any doubt. For the computation of SALI we used the \verb!LP-VI! code [\citealp{CMD14}], a fully operational routine which efficiently computes a suite of many chaos indicators for dynamical systems in any number of dimensions.

In our computations, we set $10^4$ time units as a maximum time of numerical integration. The vast majority of orbits (regular and chaotic) however, need considerable less time to find one of the two exits in the limiting curve and eventually escape from the system (obviously, the numerical integration is effectively ended when an orbit passes through one of the escape channels and escapes). Nevertheless, we decided to use such a vast integration time just to be sure that all orbits have enough time in order to escape. Remember, that there are the so called ``sticky orbits" which behave as regular ones during long periods of time. Here we should clarify, that orbits which do not escape after a numerical integration of $10^4$ time units ($10^{12}$ yr) are considered as non-escaping or trapped. In fact, orbits with escape periods equal to many Hubble times are completely irrelevant to our investigation since they lack physical meaning.

\section{Numerical results}
\label{numres}

\begin{figure*}[!tH]
\centering
\resizebox{0.8\hsize}{!}{\includegraphics{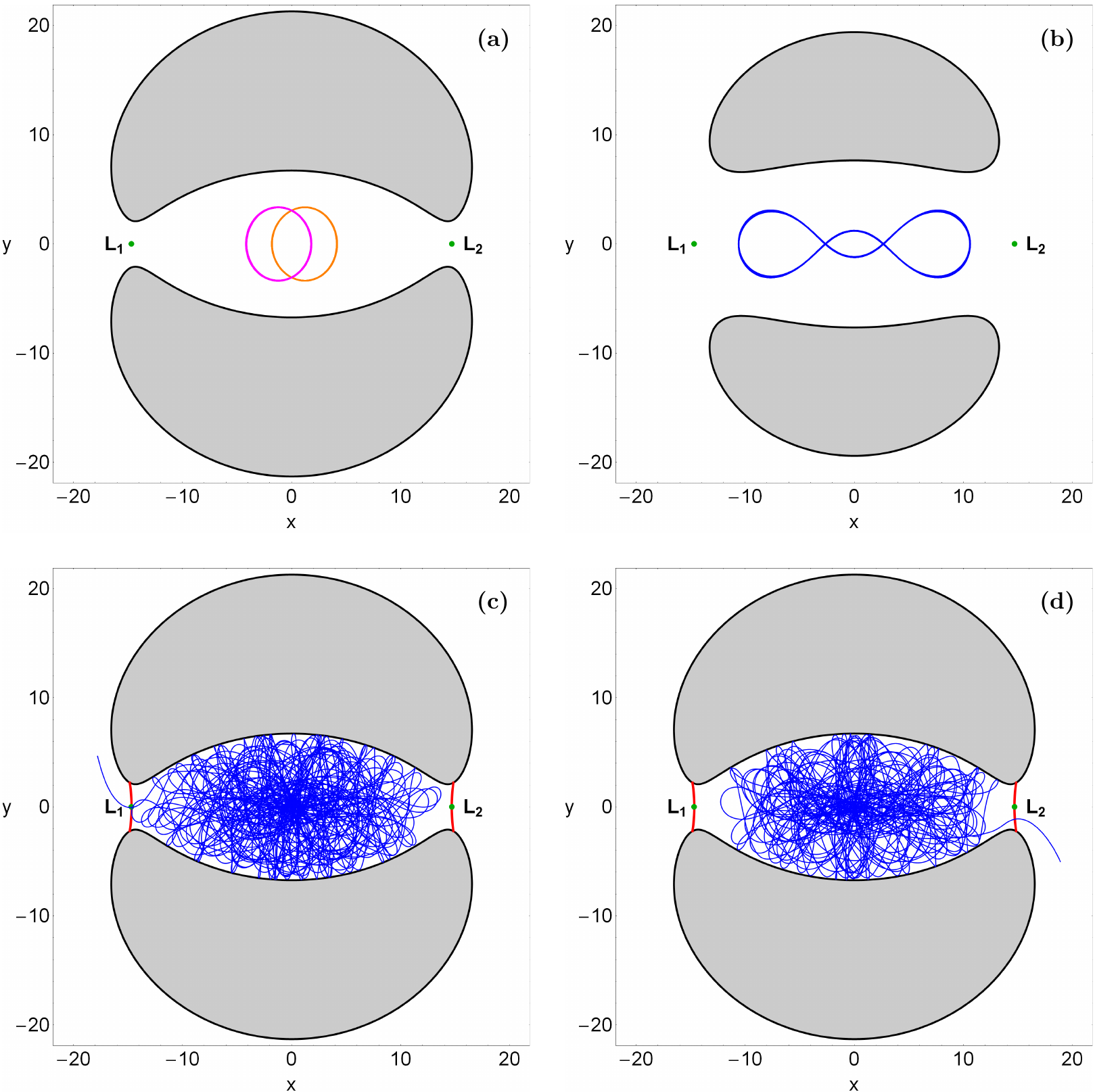}}
\caption{Characteristic examples of the main types of orbits in our galaxy model. (a-upper left): two 1:1 loop orbits, (b-upper right): a 1:3 resonant quasi-periodic box orbit, (c-lower left): an orbit escaping through $L_1$, (d-lower right): an orbit escaping through $L_2$. More details are given in Table \ref{table1}.}
\label{orbs}
\end{figure*}

\begin{table}
\centering
\setlength{\tabcolsep}{6pt}
\begin{center}
   \centering
   \caption{Type, initial conditions, value of the energy and integration time of the orbits shown in Fig. \ref{orbs}(a-d). When the initial value of $\dot{x}$ or $\dot{y}$ is indicated as $>0$ it means that is obtained from the Jacobi integral (\ref{ham}).}
   \label{table1}
   \begin{tabular}{@{}lccccccr}
      \hline
      Figure & Type & $x_0$ & $y_0$ & $\dot{x_0}$ & $\dot{y_0}$ & $\widehat{C}$ & $t_{\rm int}$  \\
      \hline
      \ref{orbs}a & 1:1 loop & -4.10 &  0.00 &   0 & $> 0$ & 0.01 & 100 \\
      \ref{orbs}a & 1:1 loop & -1.85 &  0.00 &   0 & $> 0$ & 0.01 & 100 \\
      \ref{orbs}b & 1:3 box  &  0.00 & -1.25 & $> 0$ &   0 & 0.10 & 100 \\
      \ref{orbs}c & chaotic  & -1.00 &  0.00 & $> 0$ &   0 & 0.01 & 164 \\
      \ref{orbs}d & chaotic  & -2.40 &  0.00 & $> 0$ &   0 & 0.01 & 129 \\
      \hline
   \end{tabular}
\end{center}
\end{table}

The main objective of our investigation is to determine which orbits escape and which remain trapped, distinguishing simultaneously between regular and chaotic trapped motion\footnote{Generally, any dynamical method requires a sufficient time interval of numerical integration in order to distinguish safely between ordered and chaotic motion. Therefore, if the escape rate of an orbit is very low or even worse if the orbit escapes directly from the system then, any chaos indicator (the SALI in our case) will fail to work properly due to insufficient integration time. Hence, it is pointless to speak of regular or chaotic escaping orbits.}. Furthermore, two additional properties of the orbits will be examined: (i) the channels through which the stars escape and (ii) the time-scale of the escapes (we shall also use the terms escape period or escape rates). In the present paper, we shall explore these dynamical quantities for various values of the energy, always within the interval $\widehat{C} \in [0.001,0.1]$.

\begin{figure*}[!tH]
\centering
\resizebox{\hsize}{!}{\includegraphics{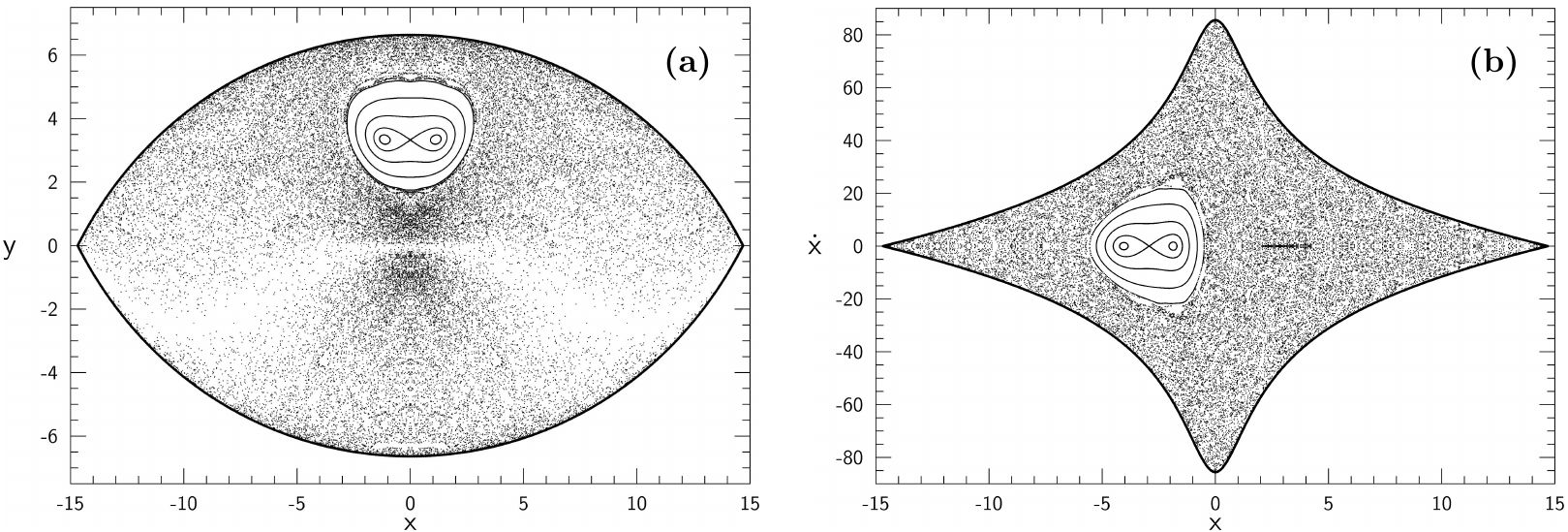}}
\caption{Poincar\'{e} Surfaces of Section at the critical Jacobi energy $C_L$. (a-left): The physical $(x,y)$ space where orbits crossing $\dot{y} = 0$ with $\dot{x} > 0$ and (b-right): the phase $(x,\dot{x})$ space where orbits crossing $y = 0$ with $\dot{y} > 0$.}
\label{pss}
\end{figure*}

Our numerical calculations indicate that apart from the escaping orbits there is always a considerable amount of non-escaping orbits. In general terms, the majority of non-escaping regions corresponds to initial conditions of regular orbits, where a third integral of motion is present, restricting their accessible phase space and therefore hinders their escape. However, there are also chaotic orbits which do not escape within the predefined interval of $10^4$ time units and remain trapped for vast periods until they eventually escape to infinity. At this point, it should be emphasized and clarified that these trapped chaotic orbits cannot be considered, by no means, neither as sticky orbits nor as super sticky orbits with sticky periods larger than $10^4$ time units. Sticky orbits are those who behave regularly for long time periods before their true chaotic nature is fully revealed. In our case on the other hand, this type of orbits exhibit chaoticity very quickly as it takes no more than about 100 time units for the SALI to cross the threshold value (SALI $\ll 10^{-8}$), thus identifying beyond any doubt their chaotic character. Therefore, we decided to classify the initial conditions of orbits in both the physical and phase space into four main categories: (i) orbits that escape through $L_1$, (ii) orbits that escape through $L_2$, (iii) non-escaping regular orbits and (iv) trapped chaotic orbits.

Additional numerical computations reveal that the non-escaping regular orbits are mainly 1:1 loop orbits for which a third integral applies\footnote{The total angular momentum is an approximately conserved quantity (integral of motion), even for orbits in non spherical potentials.}, while other types of secondary resonant orbits are also present. In Fig. \ref{orbs}a we present two 1:1 thin loop orbits, while in Fig. \ref{orbs}b a typical example of a secondary 1:3 resonant quasi-periodic orbit is shown. The $n:m$ notation we use for the regular orbits is according to [\citealp{CA98}, \citealp{ZC13}], where the ratio of those integers corresponds to the ratio of the main frequencies of the orbit, where main frequency is the frequency of greatest amplitude in each coordinate. Main amplitudes, when having a rational ratio, define the resonances of an orbit. Finally in Figs. \ref{orbs}c-d we observe two orbits escaping through $L_1$ (exit channel 1) and $L_2$ (exit channel 2), respectively. All regular orbits shown in Figs. \ref{orbs}a-b were computed until $t = 100$ time units, while on the other hand, the escaping orbits presented in Figs. \ref{orbs}c-d were calculated for 5 time units more than the corresponding escape period in order to visualize better the escape trail. In Table \ref{table1} we provide the type, the exact initial conditions and the value of the energy for all the depicted orbits.

A simple qualitative way for distinguishing between regular and chaotic motion in Hamiltonian systems is by plotting the successive intersections of the orbits using a Poincar\'{e} Surface of Section (PSS) [\citealp{HH64}]. In particular, a PSS is a two-dimensional (2D) slice of the entire four-dimensional (4D) phase space. The following Fig. \ref{pss}(a-b) shows two PSSs at the critical Jacobi energy $C_L$. Fig. \ref{pss}a corresponds to the physical $(x,y)$ space showing orbits crossing $\dot{y} = 0$ with $\dot{x} > 0$, while orbits crossing $y = 0$ with $\dot{y} > 0$ at the phase $(x,\dot{x})$ space are presented in Fig. \ref{pss}b. In both cases, the initial conditions of the orbits are integrated forwards in time plotting a dot at each crossing through the surface of section. The outermost black thick curve circumscribing each PSS is the limiting curve which in the physical space is defined as $\Phi_{\rm eff}(x,y) = E$, while in the phase space is given by
\begin{equation}
f(x,\dot{x}) = \frac{1}{2}\dot{x}^2 + \Phi_{\rm eff}(x, y=0) = E.
\label{zvc2}
\end{equation}
It is seen, that the vast majority of the surfaces of section is covered by a unified and extended chaotic sea however, stability regions with initial conditions corresponding to regular orbits are also present. In these stability islands the quasi-periodic orbits admit a third integral of motion, also known as ``adelphic integral" (see e.g., [\citealp{C63}]), which hinders the test particles (stars) from escaping. Moreover, we observe that some areas in the chaotic sea of the $(x,y)$ plane are less densely occupied than others. We remark that in the following grid exploration in the physical and phase space, we do not consider all initial conditions in the same chaotic domain as one the same orbit.

\subsection{The structure of physical $(x,y)$ space}
\label{pp1}

\begin{figure*}[!tH]
\centering
\resizebox{0.8\hsize}{!}{\includegraphics{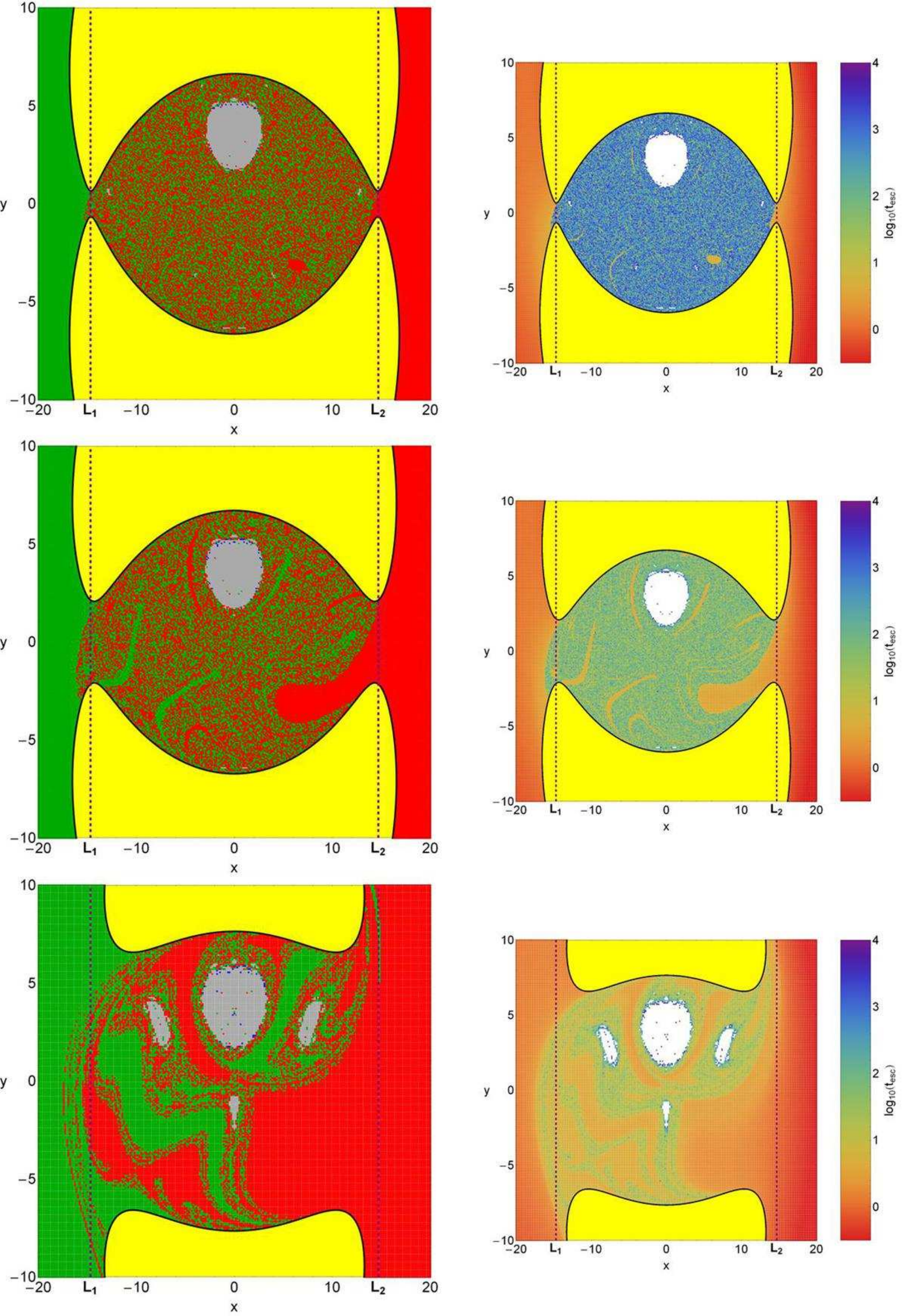}}
\caption{Left column: Orbital structure of the physical $(x,y)$ space. Top row: $\widehat{C} = 0.001$; Middle row: $\widehat{C} = 0.01$; Bottom row: $\widehat{C} = 0.1$. The green regions correspond to initial conditions of orbits where the stars escape through $L_1$, red regions denote initial conditions where the stars escape through $L_2$, gray areas represent stability islands of regular non-escaping orbits, initial conditions of trapped chaotic orbits are marked in blue, while the regions of forbidden motion are shown in yellow. The vertical, dashed, magenta lines indicate the position of the two Lagrangian points $L_1$ and $L_2$. Right column: Distribution of the corresponding escape times $t_{\rm esc}$ of the orbits on the physical space. The darker the color, the larger the escape time. The scale on the color-bar is logarithmic. Initial conditions of non-escaping regular orbits and trapped chaotic orbits are shown in white.}
\label{grd1}
\end{figure*}

Our exploration begins in the physical space and in the left column of Fig. \ref{grd1} we present the orbital structure of the $(x,y)$ plane for three values of the energy, where the initial conditions of the orbits are classified into four categories by using different colors. Specifically, gray color corresponds to regular non-escaping orbits, blue color corresponds to trapped chaotic orbits, green color corresponds to orbits escaping through channel 1, while the initial conditions of orbits escaping through exit channel 2 are marked with red color. The vertical, dashed, magenta lines indicate the position of the two Lagrangian points $L_1$ and $L_2$. For $\widehat{C} = 0.001$, that is an energy level just above the critical escape energy $C_L$, we see that the vast majority of initial conditions corresponds to escaping orbits however, a stability island of non-escaping regular orbits is present at the central upper part of the interior galactic region. We also see that the entire interior region is completely fractal which means that there is a highly dependence of the escape mechanism on the particular initial conditions of the orbits. In other words, a minor change in the initial conditions has as a result the star to escape through the opposite exit channel, which is of course, a classical indication of chaotic motion. With a much closer look at the $(x,y)$ plane we can identify some additional tiny stability islands which are deeply buried in the escape domain. As we proceed to higher energy levels the fractal area reduces and several basins of escape start to emerge. By the term basin of escape, we refer to a local set of initial conditions that corresponds to a certain escape channel. Indeed for $\widehat{C} = 0.01$ we observe the presence of escape basins which mainly have the shape of thin elongated bands. Moreover, the extent of the central stability island seems to be unaffected by the increase in the orbital energy. With increasing energy the interior region in the physical spaces becomes less and less fractal and broad, well-defined basins of escape dominate when $\widehat{C} = 0.1$. The fractal regions on the other hand, are confined mainly near the boundaries between the escape basins or in the vicinity of the stability islands. Furthermore, it is seen that for $\widehat{C} = 0.1$ apart from the central stability island, there is also a set of three smaller islands of regular motion corresponding to 1:3 resonant orbits. It should be emphasized that our classification shows that in all three energy levels trapped chaotic motion is almost negligible as the initial conditions of such orbits appear only as lonely points around the boundaries of the stability islands. Here we must point out three interesting phenomena that take place with increasing energy: (i) the regions of forbidden motion are significantly reduced, (ii) the two escape channels become more and more wide and (iii) the amount of orbits with initial conditions outside the interior region (especially outside $L_1$) that escape from the opposite exit channel increases.

\begin{figure}[!tH]
\includegraphics[width=\hsize]{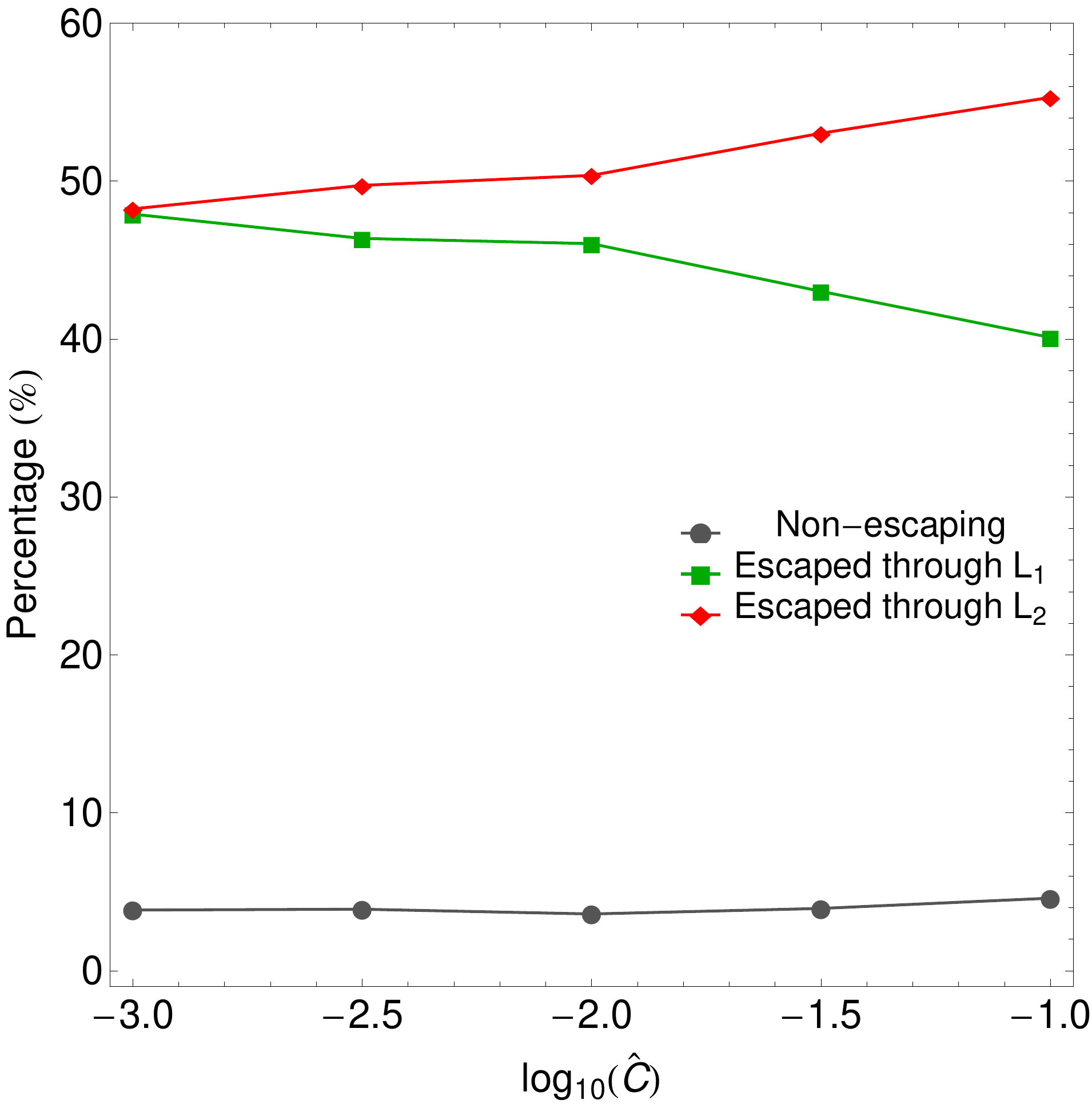}
\caption{Evolution of the percentages of escaping and non-escaping (regular plus chaotic) orbits on the physical $(x,y)$ space when varying the dimensionless energy parameter $\widehat{C}$.}
\label{percs1}
\end{figure}

In the right column of Fig. \ref{grd1} we show how the escape times $t_{\rm esc}$ of orbits are distributed on the physical $(x,y)$ space. The escape time $t_{\rm esc}$ is defined as the time when a star passes one of the two vertical, dashed lines at $(x,y) = (\pm r_L,y)$ with velocity pointing outwards. Light reddish colors correspond to fast escaping orbits with short escape periods, dark blue/purpe colors indicate large escape rates, while white color denote both non-escaping regular and trapped chaotic orbits. The scale on the color bar is logarithmic. It is evident, that orbits with initial conditions close to the boundaries of the stability islands need significant amount of time in order to escape from the galaxy, while on the other hand, inside the basins of escape where there is no dependence on the initial conditions whatsoever, we measured the shortest escape rates of the orbits. We observe that for $\widehat{C} = 0.001$ the escape periods of orbits with initial conditions in the fractal region are huge corresponding to tens of thousands of time units. As the value of the total orbital energy increases however, the escape times of orbits reduce significantly. In fact for $\widehat{C} = 0.01$ and $\widehat{C} = 0.1$ the basins of escape can be distinguished in the grids of Fig. \ref{grd1}, being the regions with reddish colors indicating extremely fast escaping orbits. Our numerical calculations indicate that orbits with initial conditions inside the basins have significantly small escape periods of less than 10 time units.

The following Fig. \ref{percs1} presents the evolution of the percentages of the different types of orbits on the physical $(x,y)$ space when the dimensionless energy parameter $\widehat{C}$ varies. At this point we must clarify that we decided to merge the percentages of non-escaping regular and trapped chaotic orbits together because our computations indicate that always the rate of trapped chaotic orbits is extremely small (less than 0.1\%) and therefore it does not contribute to the overall orbital structure of the dynamical system. One may observe, that when $\widehat{C} = 0.001$, that is just above the critical escape energy $C_L$, escaping orbits through $L_1$ and $L_2$ share about 96\% of the available physical space, so the two exit channels can be considered equiprobable. As the value of the energy increases however, the percentages of escaping orbits through exit channels 1 and 2 start to diverge, especially for $\widehat{C} > 0.01$. In particular, the amount of orbits that escape through $L_1$ exhibits a linear increase, while on the other hand, the rate of escaping orbits through the opposite exit channel displays a linear decrease. At the highest energy level studied $(\widehat{C} = 0.1)$ escaping orbits through $L_1$ is the most populated family occupying about 55\% of the physical plane, while only about 40\% of the same plane corresponds to initial conditions of orbits that escape through $L_2$. Furthermore, the percentage of non-escaping (regular plus chaotic) orbits is much less affected by the shifting of the energy displaying a rather constant value around 5\%. Thus taking into account all the above-mentioned analysis we may conclude that at low energy levels where the fractility of the physical space is maximum the stars do not show any particular preference regarding the escape channel, while on the contrary, at high enough energy levels where basins of escape dominate it seems that exit channel 2 is more preferable.

\subsection{The structure of the phase $(x,\dot{x})$ space}
\label{pp2}

\begin{figure*}[!tH]
\centering
\resizebox{0.8\hsize}{!}{\includegraphics{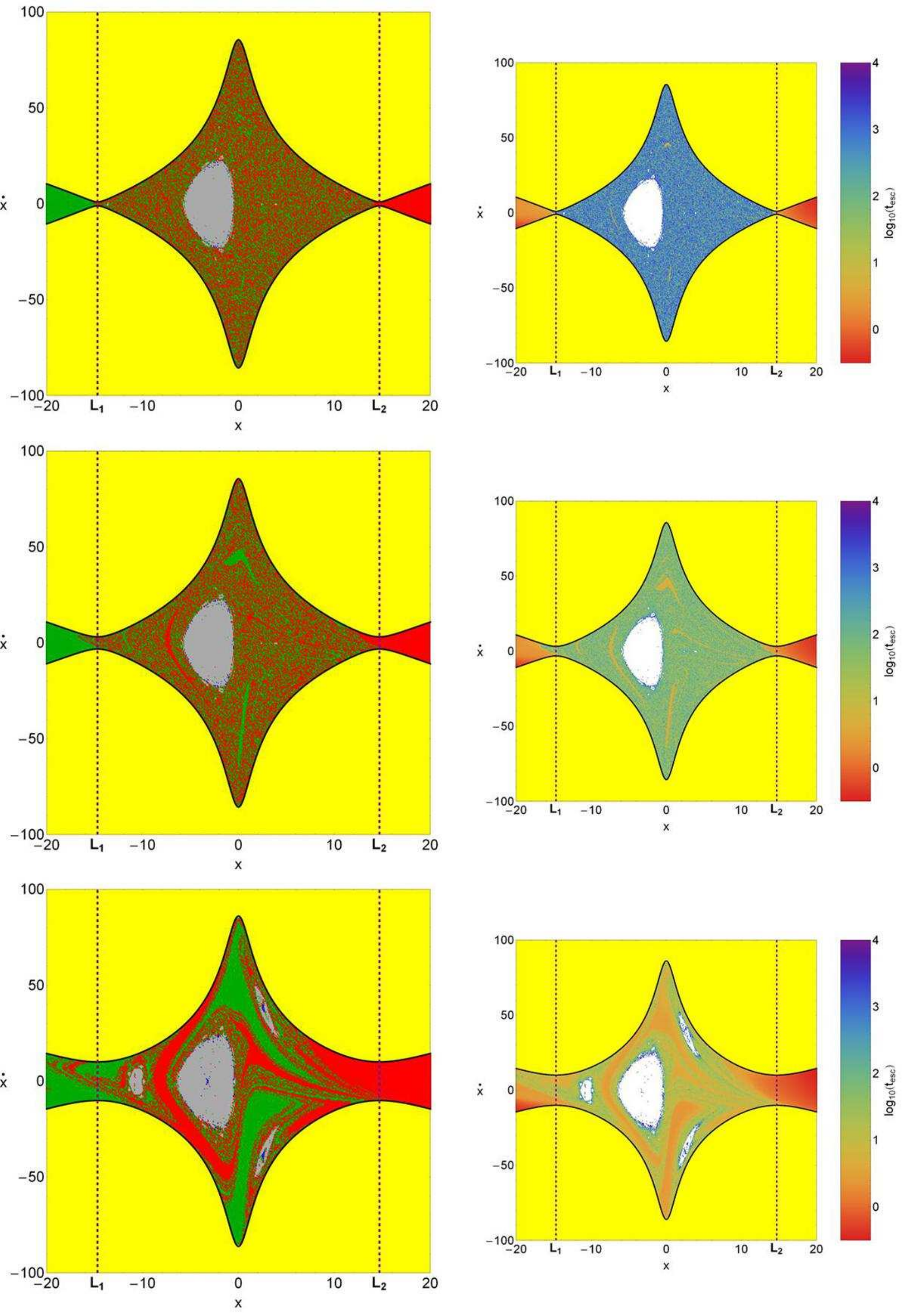}}
\caption{Left column: Orbital structure of the phase $(x,\dot{x})$ space. Top row: $\widehat{C} = 0.001$; Middle row: $\widehat{C} = 0.01$; Bottom row: $\widehat{C} = 0.1$. Right column: Distribution of the corresponding escape times $t_{\rm esc}$ of the orbits on the phase space. The color codes are the same as in Fig. \ref{grd1}.}
\label{grd2}
\end{figure*}

We continue our investigation in the phase $(x,\dot{x})$ space and we follow the same numerical approach as discussed previously. In Fig. \ref{grd2} we depict the orbital structure of the $(x,\dot{x})$ plane for three values of the energy, using different colors in order to distinguish between the four main types of orbits (non-escaping regular; trapped chaotic; escaping through $L_1$ and escaping through $L_2$). Just above the escape energy, that is when $\widehat{C} = 0.001$, we see that the structure of the phase plane is very similar to that of the corresponding physical plane shown in Fig. \ref{grd1}. It is observed that about 90\% of the $(x,\dot{x})$ plane is occupied by initial conditions of escaping orbits. Moreover, the vast escape domain is highly fractal, while basins of escape, if any, are negligible. In all studied cases the areas of regular motion correspond mainly to retrograde orbits (i.e., when a star revolves around the galaxy in the opposite sense with respect to the motion of the galaxy itself), while there are also some smaller stability islands of prograde orbits at high energy levels (i.e., for $\widehat{C} = 0.1$). The area on the phase plane covered by escape basins grows drastically with increasing energy and at high enough energy levels the fractal regions are confined mainly at the boundaries of the stability islands of non-escaping orbits. In particular, the escape basins first appear at mediocre values of energy $(\widehat{C} = 0.01)$ inside the fractal, extended escape region having the shape of thin elongated bands however, as we proceed to higher energy levels they grow in size dominating in the phase space as well-formed broad domains. It is worth noticing the initial conditions of trapped chaotic orbits at the central region of the stability islands for $\widehat{C} = 0.1$ which indicates the presence of a thin chaotic layer (separatrix\footnote{The separatrix prevents chaotic orbits from escaping in the same manner that an adelphic integral hinders quasi-periodic orbits from escaping.}). The Coriolis forces dictated by the rotation of the compact active nucleus of the Seyfert galaxy makes the phase planes to be asymmetric with respect to the $\dot{x}$-axis and this phenomenon is usually known as ``Coriolis asymmetry" (e.g., [\citealp{I80}]).

\begin{figure}[!tH]
\includegraphics[width=\hsize]{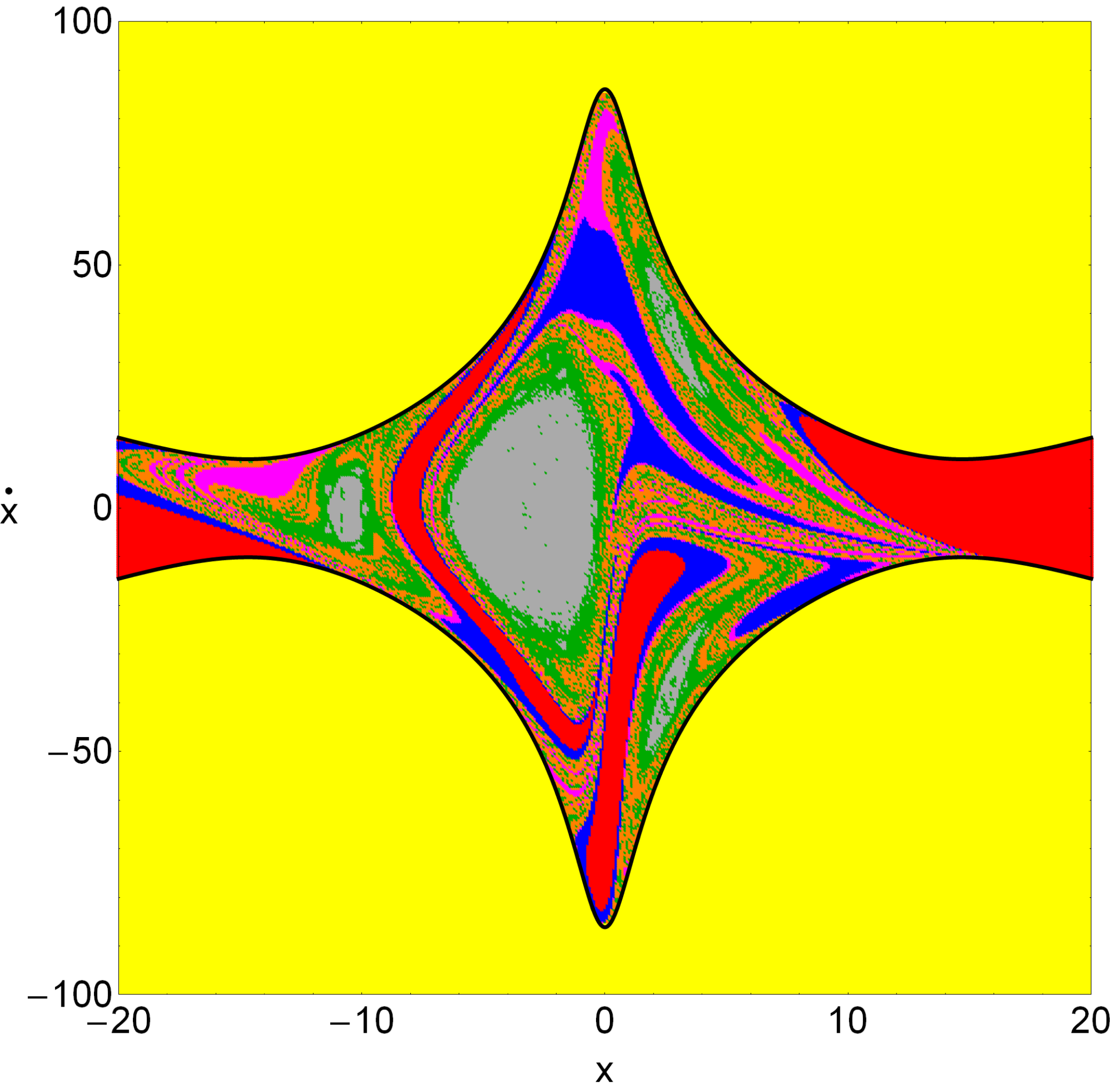}
\caption{Reconstruction of the phase $(x,\dot{x})$ plane for $\widehat{C} = 0.1$ where the color scale of the escape regions is given as a function of the number of intersections with the $y = 0$ axis upwards $(\dot{y} > 0)$. The color code is as follows: 0 intersections (red); 1 intersection (blue); 2 intersections (magenta); 3--10 intersections (orange); $> 10$ intersections (green). The gray regions represent stability islands of non-escaping orbits.}
\label{iters}
\end{figure}

\begin{figure}[!tH]
\includegraphics[width=\hsize]{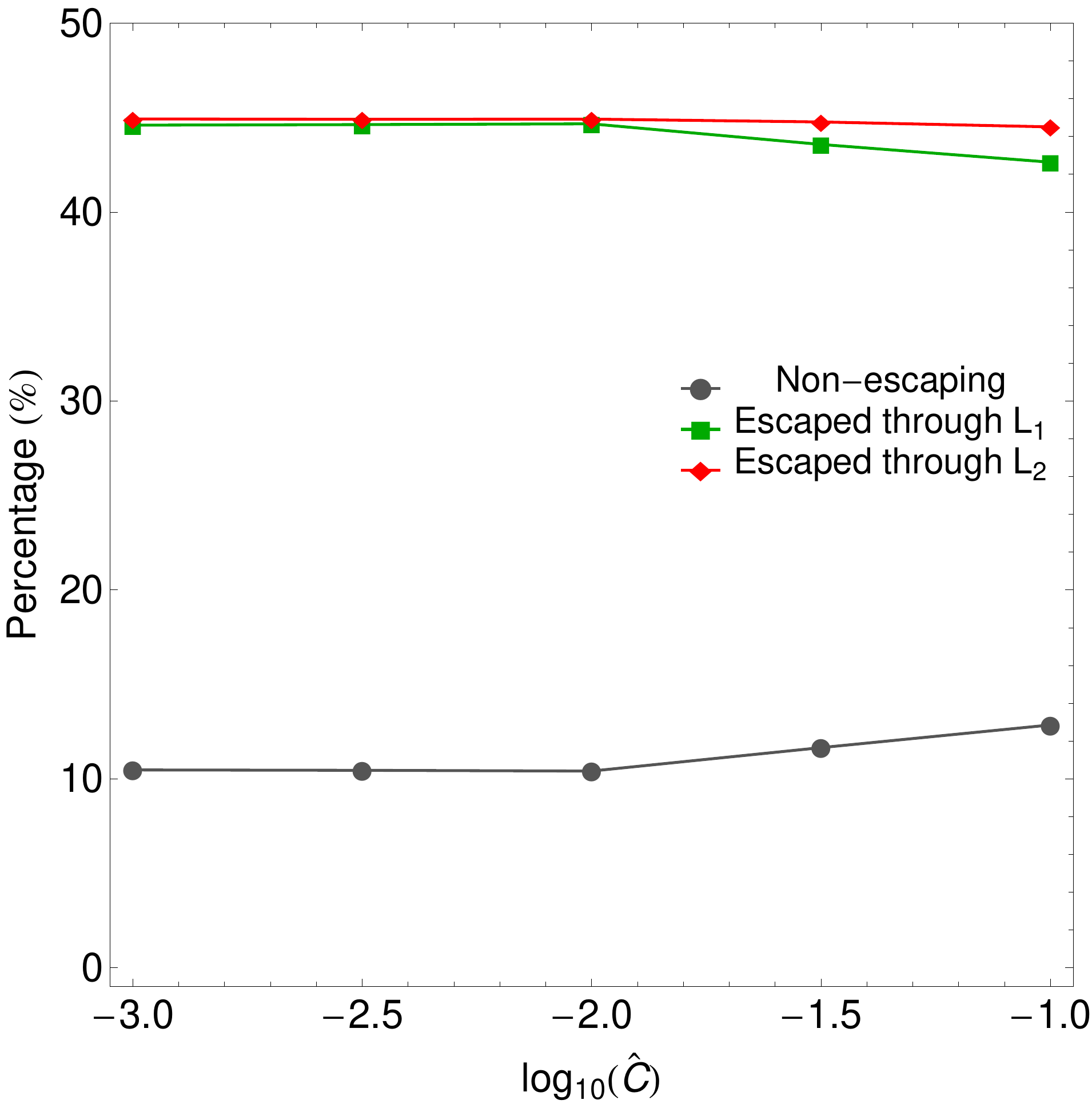}
\caption{Evolution of the percentages of escaping and non-escaping (regular plus chaotic) orbits on the phase $(x,\dot{x})$ space when varying the dimensionless energy parameter $\widehat{C}$.}
\label{percs2}
\end{figure}

The distribution of the corresponding escape times $t_{\rm esc}$ of orbits on the phase space as a function of the energy parameter is shown in the right column Fig. \ref{grd2}. One may observe that the results are very similar to those presented earlier in Fig. \ref{grd1}, where we found that orbits with initial conditions inside the basins of escape have the smallest escape rates, while on the other hand, the longest escape times correspond to orbits with initial conditions either in the fractal regions of the plots, or near the boundaries of the islands of regular motion. Another interesting way of measuring the escape period of an orbit is by counting how many intersection the orbit has with the axis $y = 0$ before it escapes. The regions of the phase space for $\widehat{C} = 0.1$ in Fig. \ref{iters} are colored according to the number of intersections with the axis $y = 0$ upwards $(\dot{y} > 0)$. Our computations reveal that orbits that escape directly without any intersection with the $y = 0$ axis (0 intersections) or performing between 3 and 10 intersections or more than 10 intersections share about three fourths of the escape region. Orbits that cross one time the $y = 0$ axis before escape correspond to about 15\% of the escape realm, while only about 10\% of the escaping orbits perform two intersections before leaving the system passing through one of the Lagrangian points. It is evident, that orbits with initial conditions located near the boundaries of the stability islands perform numerous intersections $(> 10)$ with the $y = 0$ axis before they eventually escape to infinity. On the contrary, orbits with initial conditions in the escape basins need only a couple of intersection until they escape.

The evolution of the percentages of the different types of orbits on the phase $(x,\dot{x})$ space as a function of the dimensionless energy parameter $\widehat{C}$ is presented in Fig. \ref{percs2}. Once more, we have to point out that the percentages of non-escaping ordered and trapped chaotic orbits are merged together in a single trend line because the percentage of trapped chaotic orbits is extremely small (less than 0.5\%) throughout. It is seen, that at the lowest energy level studied $(\widehat{C} = 0.001)$ escaping orbits through $L_1$ and $L_2$ share about 90\% of the phase plane, while only 10\% of the same plane corresponds to initial conditions of non-escaping (regular plus chaotic) orbits. Furthermore, we observe that for $0.001 < \widehat{C} < 0.01$ the rates of all types of orbits remain practically unperturbed by the shifting on the value of the orbital energy. For larger values of the energy $(\widehat{C} > 0.01)$ however, the percentages of escaping orbits exhibit a small divergence, while the portion of non-escaping orbits displays a minor increase and for $\widehat{C} = 0.1$ they reach about 12.5\%. Specifically, the rate of orbits escaping through $L_1$ slightly reduces, while that of orbits escaping through $L_2$ continues the monotone behavior. It should be emphasized that the divergence of the percentages of escaping orbits observed in the phase space is much smaller than that found to exist in the physical space. Therefore, one may reasonably deduce that in the phase space the increase in the value of the energy does not influence significantly the orbital content of the dynamical system, thus leaving the percentages of the orbits almost the same throughout. In addition, since that the rates of escaping orbits remain unperturbed within the energy range, we may say that the two exit channels can be considered almost equiprobable in the phase space.

\subsection{An overview analysis}
\label{over}

\begin{figure*}[!tH]
\centering
\resizebox{0.9\hsize}{!}{\includegraphics{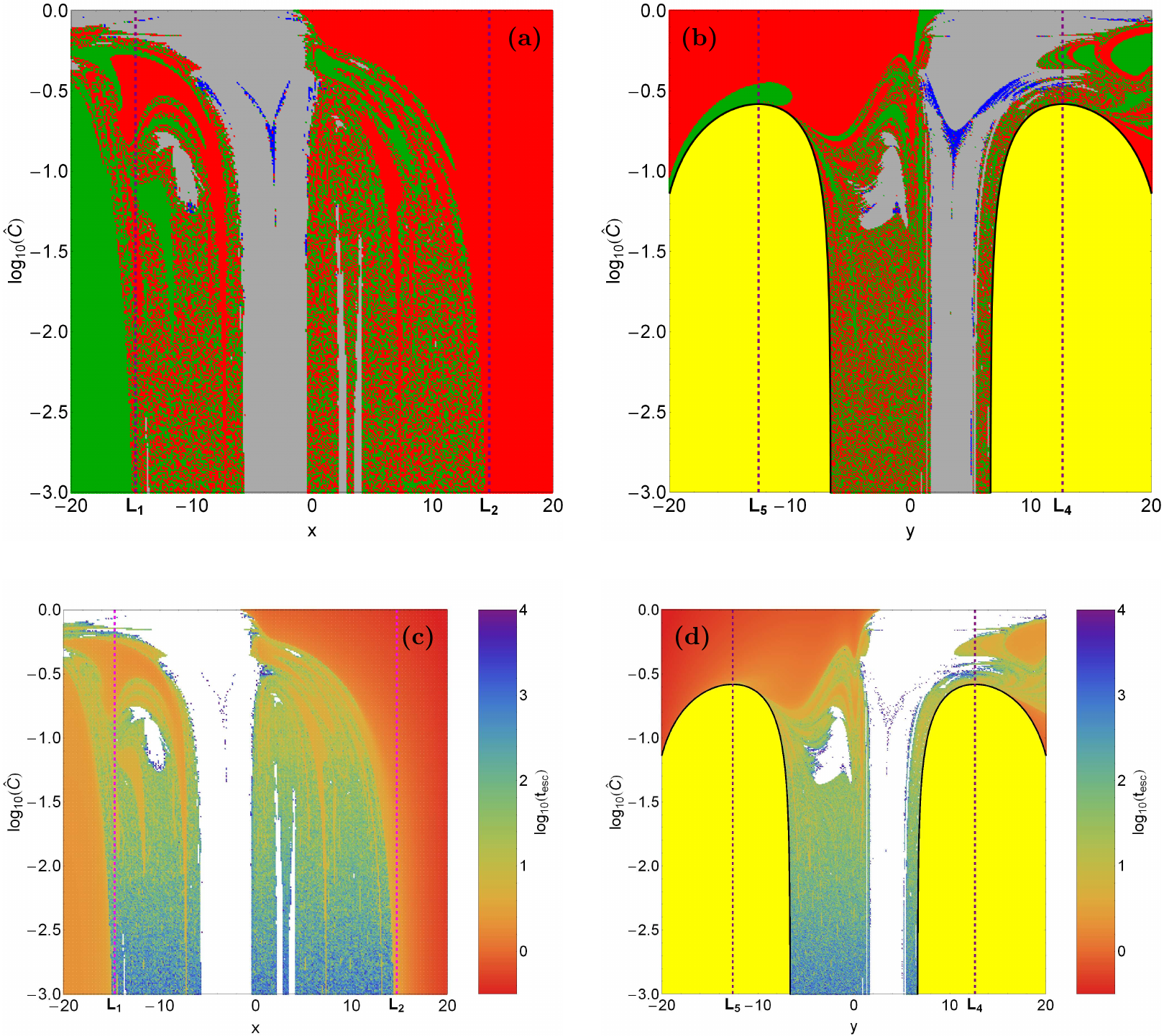}}
\caption{Orbital structure of the (a-upper left): $(x,\widehat{C})$-plane; and (b-upper right): $(y,\widehat{C})$-plane; (c-lower left and d-lower right): the distribution of the corresponding escape times of the orbits. The color codes are the exactly same as in Fig. \ref{grd1}.}
\label{xyc}
\end{figure*}

The color-coded grids in the physical $(x,y)$ as well as in the phase $(x,\dot{x})$ space provide sufficient information on the phase space mixing however, for only a fixed value of the Jacobi constant. H\'{e}non back in the late 60s [\citealp{H69}], introduced a new type of plane which can provide information not only about stability and chaotic regions but also about areas of trapped and escaping orbits using the section $y = \dot{x} = 0$, $\dot{y} > 0$ (see also [\citealp{BBS08}]). In other words, all the orbits of the stars of the galaxy are launched from the $x$-axis with $x = x_0$, parallel to the $y$-axis $(y = 0)$. Consequently, in contrast to the previously discussed types of planes, only orbits with pericenters on the $x$-axis are included and therefore, the value of the dimensionless energy parameter $\widehat{C}$ can be used as an ordinate. In this way, we can monitor how the energy influences the overall orbital structure of our dynamical system using a continuous spectrum of energy values rather than few discrete energy levels. In Fig. \ref{xyc}a we present the orbital structure of the $(x,\widehat{C})$-plane when $\widehat{C} \in [0.001,1]$, while in Fig. \ref{xyc}c the distribution of the corresponding escape times of orbits is depicted.

We observe that for relatively low energy values $(0.001 < \widehat{C} < 0.01)$ the interior galactic region is highly fractal, while basins of escape are located only at the outer parts of the $(x,\widehat{C})$-plane, that is outside $L_1$ and $L_2$ (or in other words in the exterior galactic region). Furthermore, we can identify the presence of two small stability islands of prograde $(x_0 > 0)$ non-escaping regular orbits which were invisible in the general plots shown earlier in Figs. \ref{grd1} and \ref{grd2}. For larger values of energy $\widehat{C} > 0.01$ however, the structure of the $(x,\widehat{C})$-plane changes drastically and the most important differences are the following: (i) several basins of escape are formed inside the fractal escape region, (ii) apart from the main stability island a second smaller one emerges near $L_1$. It should be pointed out that in the blow-ups of the diagram several additional extremely tiny islands of stability have been identified\footnote{An infinite number of regions of (stable) quasi-periodic (or small scale chaotic) motion is expected from classical chaos theory.}, (iii) at high enough energy levels $(\widehat{C} > 0.1)$ we see that a thin chaotic layer composed of initial conditions of trapped chaotic orbits appear at the central region of the main stability island and (iv) the extent of the main island of regular motion grows rapidly and for $\widehat{C} > 1$ the position of initial conditions of non-escaping regular orbits exceed the interior region $(x_0 > r_L)$. The following Table \ref{table2} shows the percentages of the four main types of orbits in the color-coded grids shown in Fig. \ref{xyc}(a-b).

\begin{table}[!ht]
\begin{center}
   \caption{Percentages of the four main types of orbits in the color-coded grids shown in Figs. \ref{xyc}(a-b).}
   \label{table2}
   \setlength{\tabcolsep}{6pt}
   \begin{tabular}{@{}lcccc}
      \hline
      Figure & Trapped & Non-escaping & Channel 1 & Channel 2 \\
      \hline
      \ref{xyc}a &  0.34 & 19.05 & 31.94 & 48.67 \\
      \hline
      \ref{xyc}b &  1.46 & 28.72 & 26.53 & 43.29 \\
      \hline
   \end{tabular}
\end{center}
\end{table}

In order to obtain a more complete view on how the total orbital energy influences the nature of orbits in our galaxy model, we follow a similar numerical approach to that explained before but now all orbits of stars are initiated from the vertical $y$-axis with $y = y_0$. In particular, this time we use the section $x = \dot{y} = 0$, $\dot{x} > 0$, launching orbits parallel to the $x$-axis. This allow us to construct again a 2D plane in which the $y$ coordinate of orbits is the abscissa, while the logarithmic value of the energy $\log_{10}(\widehat{C})$ is the ordinate. The orbital structure of the $(y,\widehat{C})$-plane when $\widehat{C} \in [0.001,1]$ is shown in Fig. \ref{xyc}b. The vertical, dashed, magenta lines indicate the position of the Lagrangian points $L_4$ and $L_5$. The black solid line is the limiting curve which distinguishes between regions of allowed and forbidden motion and is defined as
\begin{equation}
f_L(y,\widehat{C}) = \Phi_{\rm eff}(0,y) = E.
\label{zvc2}
\end{equation}
A very complicated orbital structure is reveled in the $(y,\widehat{C})$-plane which however, in general terms, is very similar with that of the $(x,\widehat{C})$-plane. One may observe that for $E > -374.55$ the forbidden regions of motion around $L_4$ and $L_5$ completely disappear. In fact the value $E(L_4) = E(L_5) = -374.55$ is another critical energy level. It is seen that above that critical value the left part $(y_0 < 0)$ of the plane is occupied almost completely with initial conditions of orbits that escape through $L_2$, while the right part of the same plane on the other hand, contains mostly initial conditions of non-escaping regular orbits.

\begin{figure*}[!tH]
\centering
\resizebox{0.9\hsize}{!}{\includegraphics{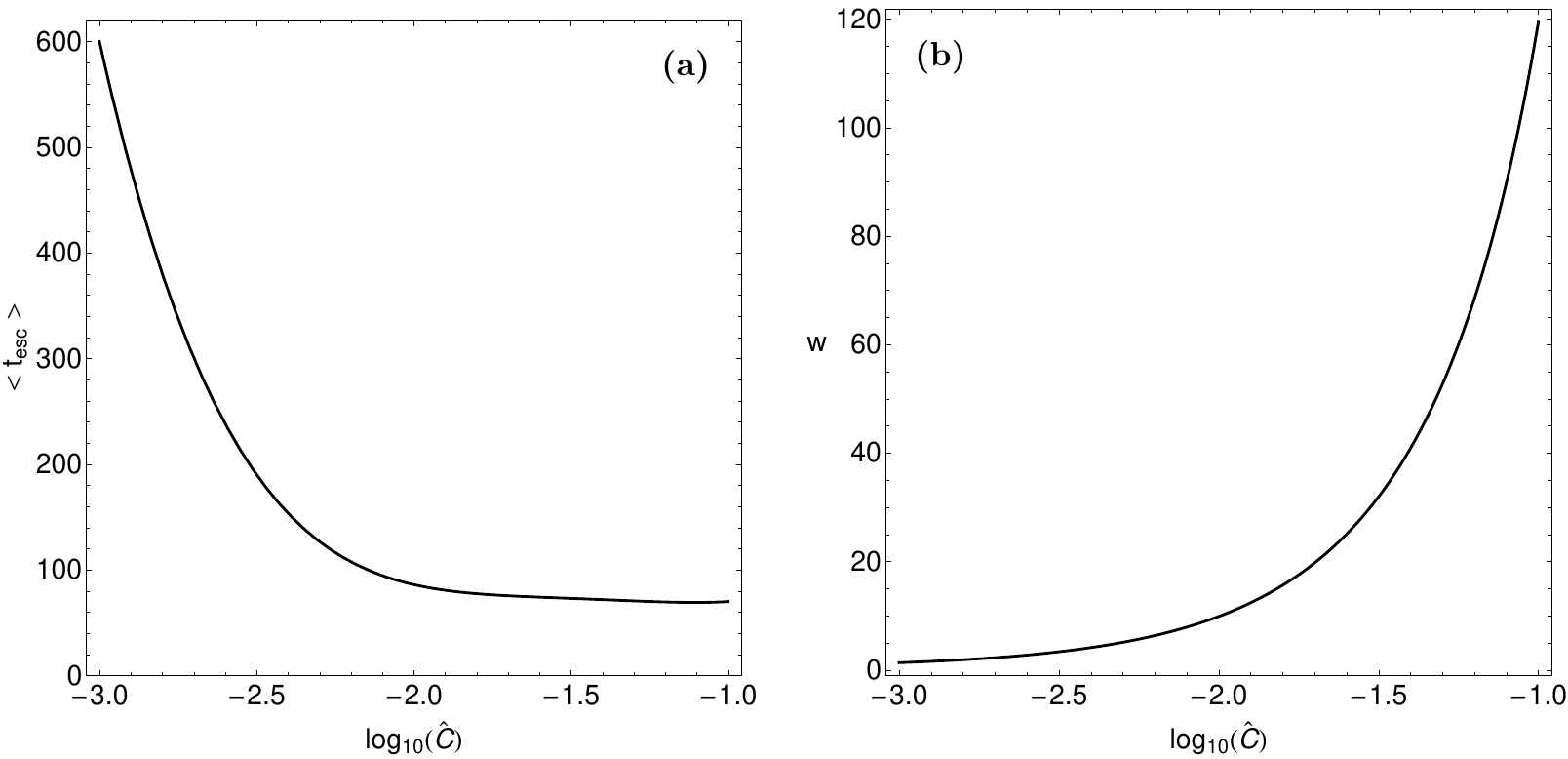}}
\caption{Evolution of (a-left): the average escape time of orbits $< t_{\rm esc} >$ and (b-right): the width $w$ of the escape channels as a function of the orbital energy $\widehat{C}$.}
\label{wtesc}
\end{figure*}

It is evident from the results presented in Figs. \ref{xyc}(c-d) that the escape times of the orbits are strongly correlated to the escape basins. In addition, one may conclude that the smallest escape periods correspond to orbits with initial conditions inside the escape basins, while orbits initiated in the fractal regions of the planes or near the boundaries of stability islands have the highest escape rates. In both types of planes the escape times of orbits are significantly reduced with increasing energy. Combining all the numerical outcomes presented in Figs. \ref{xyc}(a-d) we may say that the key factor that determines and controls the escape times of the orbits is the value of the orbital energy (the higher the energy level the shorter the escape rates), while the fractality of the basin boundaries varies strongly both as a function of the energy and of the spatial variable.

The evolution of the average value of the escape time $< t_{\rm esc} >$ of orbits as a function of the dimensionless energy parameter is given in Fig. \ref{wtesc}a. It is seen, that for low values of energy the average escape time of orbits reduces rapidly, while for $\widehat{C} > 0.01$, it seems to saturate around 80 time units. We feel it is important to justify this behaviour of the escape time. As previously explained, a Lyapunov orbit acts like a barrier, reaches the ZVC on both sides of the opening and returns along the same path thus, connecting two opposite branches of the ZVC. Therefore, we could exploit this fact in order to quantify the escape process. For this purpose, we define $w$ as the distance between the two points at which the Lyapunov orbit touches the two opposite branches of the ZVC. Using $w$ we can measure, in a way, the width of the escape channels. Fig. \ref{wtesc}b depicts the evolution of $w$ as a function of the energy parameter $\widehat{C}$. We observe an almost exponential growth of $w$ with increasing energy. This means that as the value of the Jacobi constant increases the escape channels (which are of course symmetrical) become more and more wide and therefore, orbits need less and less time in order to find one of the two openings in the open equipotential surface and eventually escape from the system. This geometrical feature explains why for low values of energy orbits consume large time periods wandering inside the open equipotential surface until they eventually locate one of the two exits and escape to infinity. Therefore, we may conclude that it is the geometrical shape of the exit channels that justifies the decreasing pattern of $t_{\rm esc}$ shown in Fig. \ref{wtesc}a.

\begin{figure}[!tH]
\includegraphics[width=\hsize]{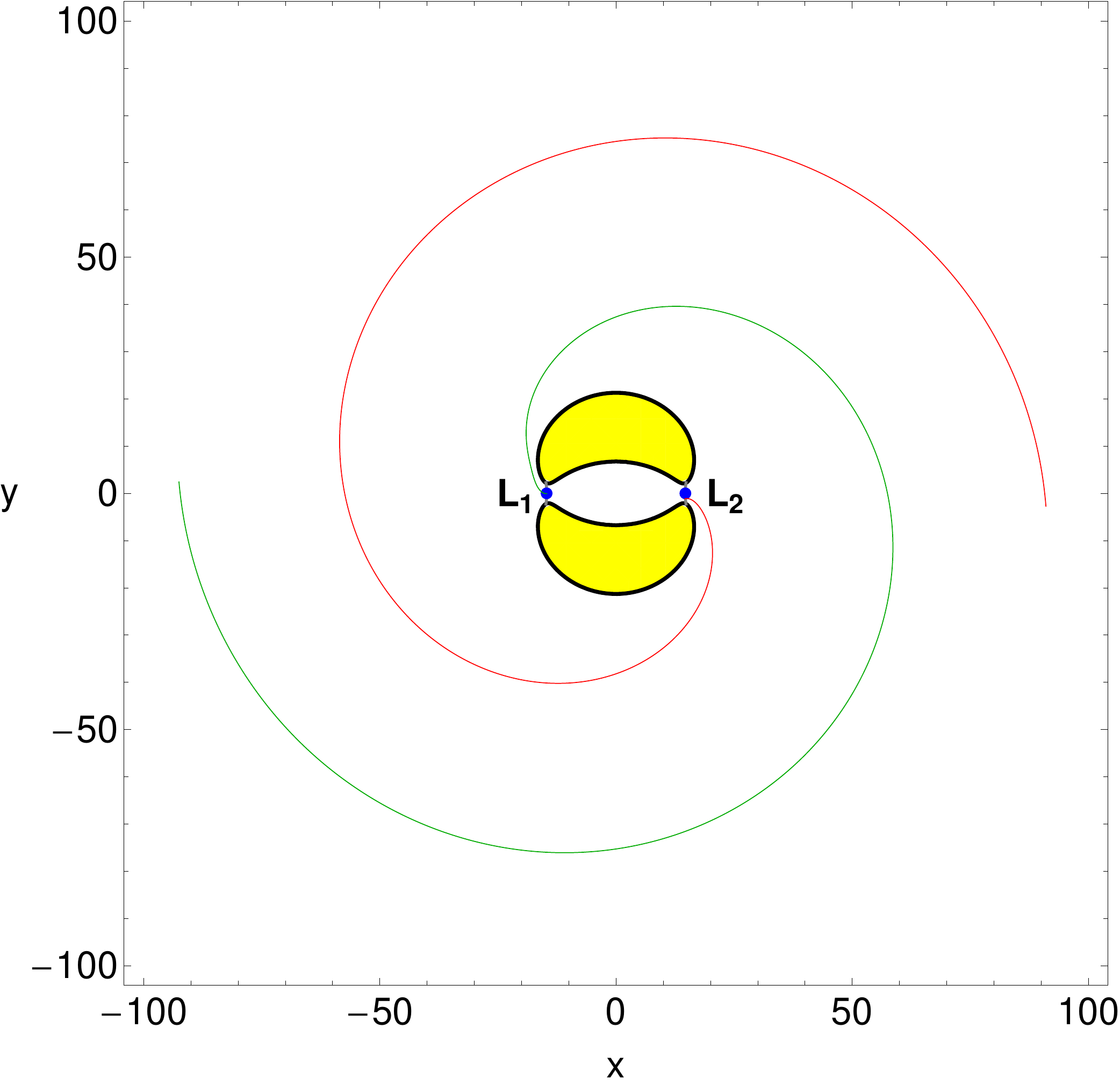}
\caption{Visualization of the spiral path of two orbits after escaping from $L_1$ (green) and $L_2$ (red) when $\widehat{C} = 0.01$. The blue dots indicate the position of $L_1$ and $L_2$, while the yellow banana-shaped isolines define the forbidden regions of motion around $L_4$ and $L_5$.}
\label{sp0}
\end{figure}

\begin{figure*}[!tH]
\centering
\resizebox{0.8\hsize}{!}{\includegraphics{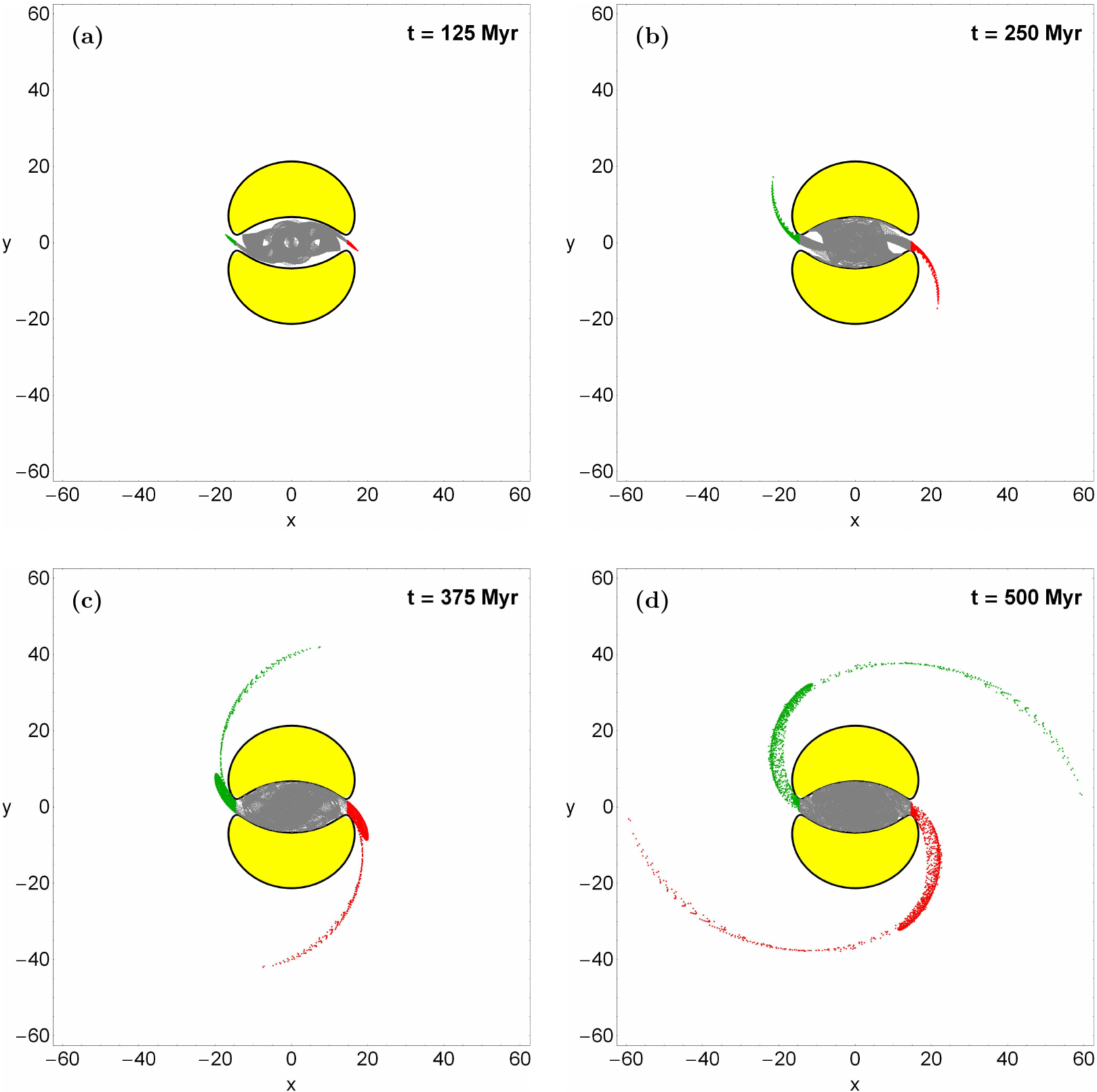}}
\caption{Simulation time snapshots of the position of $10^6$ stars in the physical $(x,y)$ plane initiated $(t = 0)$ within the Lagrangian radius $r_L$, for $\widehat{C} = 0.01$. We see that as time evolves two symmetrical spiral arms are formed. The green arm contains stars that escaped through $L_1$, while the red arm contains stars that escaped through $L_2$. The bounded (trapped or non-escaping) stars inside the interior galactic region are shown in gray.}
\label{sps}
\end{figure*}

\section{Formation of spiral arms}
\label{spr}

It is well known that when stars escape from star clusters through the Lagrangian points form complicated structures known as tidal tails or tidal arms (e.g., [\citealp{CMM05}, \citealp{dMCM05}, \citealp{JBPE09}, \citealp{KMH08}, \citealp{KKBH10}]). In the same vein, in the case of barred spiral galaxies we observe the formation of spiral arms (e.g., [\citealp{EP14}, \citealp{GKC12}, \citealp{MT97}, \citealp{MQ06}, \citealp{QDB11}, \citealp{RDQ08}, \citealp{SK91}]). Usually a star cluster rotates around its parent galaxy in a circular orbit thus, the tidal tails follow the curvature defined by the circular orbit and they are nearly straights for orbits moving in large galactocentric distances (see e.g., [\citealp{JBPE09}]). The spiral arms of barred galaxies on the other hand, are formed by the non-axisymmetric perturbation of the bar. In particular, in barred spiral galaxies with prominent spiral-like shape the two arms start from the ends of the bar and then they wind up around the banana-shaped forbidden regions which enclose the Lagrangian points $L_4$ and $L_5$.

Almost all Seyfert galaxies are spiral galaxies and have been among the most intensively studied objects in astronomy (e.g., [\citealp{SSN14}]). Therefore, it would be very challenging to investigate if our simple gravitational model has the ability to realistically simulate the formation of spiral arms. As a first step, we integrated the two orbits shown in Fig. \ref{orbs}(c-d) for 7 time units after crossing the Lagrangian points and in Fig. \ref{sp0} we present the corresponding paths of the orbits that escape from opposite exits. It is seen, that both orbits follow indeed a spiral path around the yellow banana-shaped isolines of the effective potential in which the Lagrangian points $L_4$ and $L_5$ are enclosed.

Taking into account that our initial tests are positive regarding the formation of the spiral structure, we reconfigured our integration routine for the computation of the basins of escape to yield output of all orbit trajectories for a three-dimensional grid of size $N_x \times N_y \times N_{\dot{x}} = 100 \times 100 \times 100$ for $\widehat{C} = 0.01$. This dense grid of initial conditions is centered at the origin of the phase space and we allow for all orbits both sings of $\dot{y}$. Our purpose is to monitor the time-evolution of the path of the orbits and simulate the spiral arm formation. At $t = 0$ all orbits are regularly distributed within the Lagrangian radius inside the interior galactic region. In Fig. \ref{sps} we present four time snapshots of the position of the $10^{6}$ stars in the physical $(x,y)$ plane. We observe that at $t = 1.25$ time units, which corresponds to 125 Myr the majority of stars are still inside the interior galactic region however, a small portion of stars has already escaped passing through either $L_1$ or $L_2$. At $t = 2.5$ time units (250 Myr) we have the first indication of the formation of spiral arms, where with green and red color we depict stars that escaped through exit channels 1 and 2, respectively, while stars that remain inside the Lagrangian radius are shown in gray. As time goes by we see that the two symmetrical spiral arms grow in size and at $t = 5$ time units (500 Myr) the Seyfert galaxy has obtained its complete spiral structure, while the interior galactic region is uniformly filled. At this point we have to point out that the standard integration code, which was used to obtain the results presented in Figs. \ref{grd1} and \ref{grd2}, plots a point at every step of the numerical integration. In Fig. \ref{sps}(a-d) on the other hand, the density of points along one star orbit is taken to be proportional to the velocity of the star. Being more specific, a point is plotted (showing the position of a star), if an integer counter variable which is increased by one at every integration step, exceeds the velocity of the star. Following this technique we can simulate, in a way, a real $N$-body simulation of the spiral evolution of the galaxy, where the density of stars will be highest where the corresponding velocity is lowest. Therefore, we proved that for $\widehat{C} = 0.01$ the scenario of evolution of spiral arms in our galaxy model is indeed viable. Additional numerical simulations reveal that this is also true for other (lower or higher) values of energy. In fact, the particular energy level mainly controls how tight the spiral arms wind up around the forbidden regions of motion.

\begin{figure}[!tH]
\includegraphics[width=\hsize]{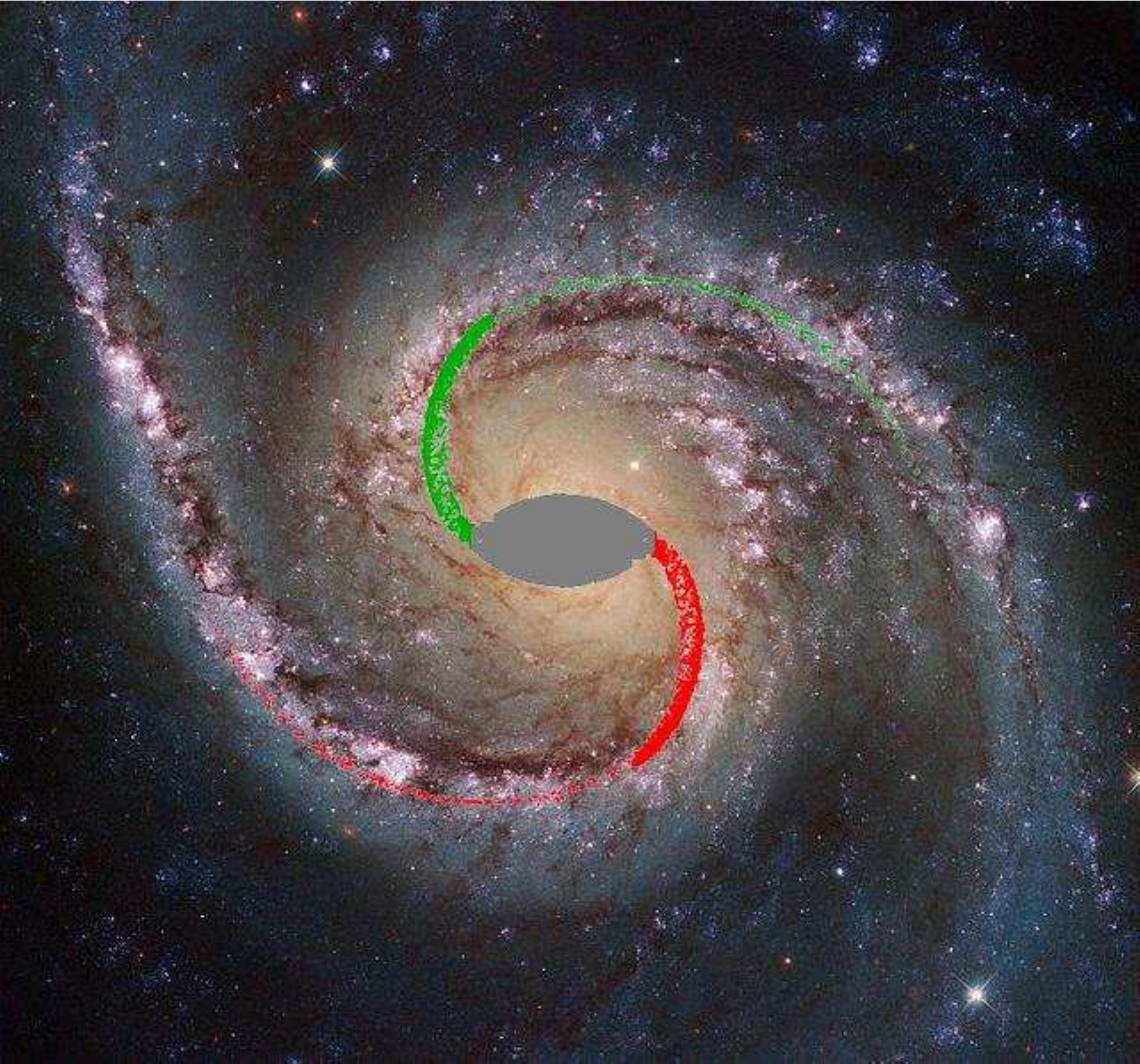}
\caption{A real image of the Seyfert galaxy NGC 1566 where we have superposed the simulation output of our gravitational model at $t$ = 500 Myr shown in Fig. \ref{sps}d. The gravitational potential studied in this paper is overall potential of the galaxy.}
\label{ngc}
\end{figure}

In section \ref{mod} we explained that the values of the mass $M_{\rm g}$ and the scale length $c$ entering Eq. (\ref{pot}) were chosen having in mind the Seyfert galaxy NGC 1566 with total mass equal to 1.2 $\times 10^{11}$ ${\rm M}_\odot$ [\citealp{KKB05}]. Here we have to point out that NGC 1566 is not just any Seyfert galaxy; it is the second brightest Seyfert galaxy known. It is also the brightest and most dominant member of the Dorado Group, a loose concentration of galaxies located approximately 40 million light-years away that together comprise one of the richest galaxy groups of the southern hemisphere. NGC 1566 is an intermediate spiral galaxy of type SAB(rs)bc, meaning that while it does not have a well defined bar-shaped region of stars at its central region - like normal barred spirals - it is not quite an unbarred spiral either. The small but extremely bright nucleus of NGC 1566 is clearly visible in Fig. \ref{ngc} [\citealp{img}], a telltale sign of its membership of the Seyfert class of galaxies. The centres of such galaxies are very active and luminous, emitting strong bursts of radiation and potentially harbouring super-massive black holes that are many millions of times the mass of the Sun. This image highlights the beauty and awe-inspiring nature of this unique galaxy group, with NGC 1566 glittering and glowing, its bright nucleus framed by swirling and symmetrical lavender arms. In Fig. \ref{ngc} we have superposed the time snapshot at $t = 5$ time units (500 Myr) of our simulation above the real image of the galaxy. We observe that the central interior region adequately matches the active galactic core, while the two spiral arms obtained from the numerical simulation fit \comm{exactly} the corresponding real ones. Thus, \comm{it is evident and beyond any doubt,} we may conclude that our simple one-component gravitational effective potential (\ref{eff}) can, in a way, \comm{realistically} model not only the escape dynamics but also the spiral formation and evolution of Seyfert galaxies.

Undoubtedly, a Seyfert galaxy is a very complex stellar entity and therefore, we need to assume some necessary simplifications and assumptions in order to be able to study mathematically the orbital behavior of such a complicated stellar system. For this purpose, our gravitational model is intentionally simple and contrived, in order to give us the ability to study all the different aspects regarding the kinematics and dynamics of the system. Nevertheless, contrived models can surely provide an insight into more realistic stellar systems, which unfortunately are very difficult to be explored, if we take into account all the astrophysical aspects (i.e., gas, mergers, close encounters etc). Self-consistent models on the other hand, are mainly used when conducting $N$-body simulations. However, this application is entirely out of the scope of the present paper. Once more, we have to point out that the simplicity of our model is necessary, otherwise it would be extremely difficult, or even impossible, to apply the extensive and detailed numerical calculations presented in this study. Our single-component potential neglect many features which must exist in real Seyfert galaxies such as colors/ages of stars populations, deviations from simple symmetries and time-independent perturbations. However, it is still significant that the conclusions derived from this simple potential all agree, both qualitatively and semi-quantitatively. This suggests that our model is robust, depending only on such topological properties of the phase space.

\section{Discussion and conclusions}
\label{disc}

The aim of this work was to shed some light to the trapped or escaping nature of orbits in a dynamical gravitational model describing a Seyfert galaxy. In order to describe the properties of the Seyfert galaxy as a whole we used an asymmetric Plummer-type potential. At this point we should emphasize that a Seyfert galaxy is composed of a disk, bulge a central nucleus and a bar. However, our dynamical system has two degrees of freedom and for this type of Hamiltonians we usually adopt single component potentials to model the properties of barred galaxies (e.g., [\citealp{BT08}, \citealp{C00}, \citealp{CK98}, \citealp{C81}, \citealp{C83}, \citealp{CP80}]) or galaxies in general (e.g., [\citealp{BBP06}, \citealp{CP03a}, \citealp{CP03b}, \citealp{CM85}, \citealp{KC01}, \citealp{KC02}, \citealp{MAB95}, \citealp{VWD12}]). This approach allow us to directly determine the effects of the single component potential to the the different types (resonances) of orbits, the amount of chaos, the escape properties and other dynamical aspects.

The Jacobi integral of motion has a critical threshold value (or escape energy) which determines if the escape process is possible or not. In particular, for energies smaller than the threshold value, the equipotential surface is closed and it is certainly true that escape is impossible. For energy levels larger than the escape energy however, the equipotential surface opens and two exit channels appear through which the particles can escape to infinity. Here it should be emphasized, that if a test particle does has energy larger than the escape value, there is no guarantee that the star will certainly escape from the system and even if escape does occur, the time required for an orbit to transit through an exit and hence escape to infinity may be vary long compared with the natural escaping time. We managed to distinguish between ordered/chaotic and trapped/escaping orbits and we also located the basins of escape leading to different exit channels, finding correlations with the corresponding escape times of the orbits. Our extensive and thorough numerical investigation strongly suggests, that the overall escape mechanism is a very complicated procedure and very dependent on the value of the Jacobi integral.

Since a distribution function of the model was not available so as to use it for extracting the different samples of orbits, we had to follow an alternative path. We defined for several values of the Jacobi integral, dense grids of initial conditions $(x_0,y_0)$ and $(x_0, \dot{x_0})$ regularly distributed in the area allowed by the energy level on the physical and phase space, respectively and then we identified regions of order/chaos and bound/escape. In each type of grid the step separation of the initial conditions along the axes (or in other words the density of the grid) was controlled in such a way that there are always at least $10^5$ orbits to be integrated and classified. For the numerical integration of the orbits in each type of grid, we needed about between 1 hour and 5 days of CPU time on a Pentium Dual-Core 2.2 GHz PC, depending on the escape rates of orbits in each case. For each initial condition, the maximum time of the numerical integration was set to be equal to $10^4$ time units however, when a particle escaped the numerical integration was effectively ended and proceeded to the next available initial condition.

The present article provides quantitative information regarding the escape dynamics in spiral Seyfert galaxies. The main numerical results of our research can be summarized as follows:
\begin{enumerate}
 \item In all examined cases, areas of bounded motion and regions of initial conditions leading to escape in a given direction (basins of escape), were found to exist in both the physical and the phase space. The several escape basins are very intricately interwoven and they appear either as well-defined broad regions or thin elongated bands. Regions of bounded orbits first and foremost correspond to stability islands of regular orbits where a third integral of motion is present.
 \item A strong correlation between the extent of the basins of escape and the value of the Jacobi integral was found to exist. Indeed, for low energy levels the structure of both the physical and the phase space exhibits a large degree of fractalization and therefore the majority of orbits escape choosing randomly exit channels. As the value of the energy increases however, the structure becomes less and less fractal and several basins of escape emerge. The extent of these basins of escape is more prominent at high energy levels.
 \item In both the physical and the phase space we identified a small portion of trapped chaotic orbits which do not escape within the predefined maximum time of numerical integration. The initial conditions of these orbits are located either in thin chaotic layers (separatrix) inside the regular areas of motion, or near the boundaries of stability islands. Moreover, these orbits reveal their chaotic nature very quickly, so they cannot be considered as sticky orbits.
 \item We observed, that in many cases the escape process is highly sensitive dependent on the initial conditions, which means that a minor change in the initial conditions of an orbit lead the test particle to escape through the other exit channel. These regions are the exact opposite of the escape basins, are completely intertwined with respect to each other (fractal structure) and are mainly located in the vicinity of stability islands. This sensitivity towards slight changes in the initial conditions in the fractal regions implies, that it is impossible to predict through which exit the particle will escape.
 \item Our calculations revealed, that the escape times of orbits are directly linked to the basins of escape. In particular, inside the basins of escape as well as relatively away from the fractal domains, the shortest escape rates of the orbits had been measured. On the other hand, the longest escape periods correspond to initial conditions of orbits either near the boundaries between the escape basins or in the vicinity of the stability islands.
 \item It was detected, that for energy levels slightly above the escape energy the majority of the escaping orbits have considerable long escape rates (or escape periods), while as we proceed to higher energies the proportion of fast escaping orbits increases significantly. This behavior was justified, by taken into account a geometrical feature of the equipotential surface. Specifically, for low values of energy the width of the exit channels is too small therefore, stellar orbits consume large time periods wandering inside the open equipotential surface until they eventually locate one of the two exits and escape to infinity. Moreover, it was proved that the width of the escape channels grows rapidly with increasing energy, while at the same time the average escape time decreases significantly.
 \item We presented evidence that escaping stars with values of energy higher than the critical Jacobi value form spiral arms outside the interior galactic region defined by the Lagrangian radius. This procedure has a striking similarity to the formation of tidal tails in star clusters. We also conducted numerical simulations proving that our gravitational effective potential \comm{is realistic taking into account that it} can \comm{sufficiently} model the formation as well as the evolution of spiral arms in real Seyfert galaxies.
\end{enumerate}

Judging by the detailed and novel outcomes we may say that our task has been successfully completed. We hope that the present numerical analysis and the corresponding results to be useful in the field of escape dynamics in Active Galactic Nuclei (AGN). The outcomes as well as the conclusions of the present research are considered, as an initial effort and also as a promising step in the task of understanding the escape mechanism of orbits in AGNs. Taking into account that our results are encouraging, it is in our future plans to modify properly our galactic model in order to expand our investigation into three dimensions and explore the entire six-dimensional phase space. In addition, $N$-body simulations may elaborate the formation as well as the evolution of spiral structure in Seyfert galaxies.

\section*{Acknowledgments}

The author would like to thank the two anonymous referees for the careful reading of the manuscript and for all the apt suggestions and comments which allowed us to improve both the quality and the clarity of the paper.

\end{document}